\documentclass[
preprint,
 amsmath,amssymb,
prb,
]{revtex4}

\usepackage{graphicx}
\usepackage{dcolumn}
\usepackage{bm}
\usepackage{hyperref}
\usepackage[mathlines]{lineno}
\usepackage[utf8]{inputenc}
\usepackage[T1]{fontenc}
\usepackage{mathptmx}
\usepackage{xcolor}
\usepackage{array}
\usepackage{longtable}
\usepackage{adjustbox}

\begin{document}

\title{Anomalous Hall effect in 3\textit{d}/5\textit{d} multilayers mediated by interface scatterings and the nonlocal spin-conductivity.}
\author{T. H. Dang}
\affiliation{Unit\'e Mixte de Physique CNRS/Thales, University Paris-Sud and Universit\'e Paris-Saclay, 91767 Palaiseau, France}
\author{Q. Barbedienne}
\affiliation{Unit\'e Mixte de Physique CNRS/Thales, University Paris-Sud and Universit\'e Paris-Saclay, 91767 Palaiseau, France}
\author{D.~Q. To}
\affiliation{Unit\'e Mixte de Physique CNRS/Thales, University Paris-Sud and Universit\'e Paris-Saclay, 91767 Palaiseau, France}
\author{E. Rongione}
\affiliation{Unit\'e Mixte de Physique CNRS/Thales, University Paris-Sud and Universit\'e Paris-Saclay, 91767 Palaiseau, France}
\author{N. Reyren}
\affiliation{Unit\'e Mixte de Physique CNRS/Thales, University Paris-Sud and Universit\'e Paris-Saclay, 91767 Palaiseau, France}
\author{F. Godel}
\affiliation{Unit\'e Mixte de Physique CNRS/Thales, University Paris-Sud and Universit\'e Paris-Saclay, 91767 Palaiseau, France}
\author{S. Collin}
\affiliation{Unit\'e Mixte de Physique CNRS/Thales, University Paris-Sud and Universit\'e Paris-Saclay, 91767 Palaiseau, France}
\author{J.~M. George}
\affiliation{Unit\'e Mixte de Physique CNRS/Thales, University Paris-Sud and Universit\'e Paris-Saclay, 91767 Palaiseau, France}
\author{H. Jaffr\`es}
\date{\today}
\affiliation{Unit\'e Mixte de Physique CNRS/Thales, University Paris-Sud and Universit\'e Paris-Saclay, 91767 Palaiseau, France}

\begin{abstract}
We have evidenced unconventional Anomalous Hall Effects (AHE) in \textit{3d}/\textit{5d} (Co0.2nm/Ni0.6nm)$_N$ multilayers grown on a thin Pt layer or thin Au:W alloys with perpendicular magnetic anisotropy (PMA) properties. The inversion of AHE observed with one Pt series is explained by considering opposite sign of the effective spin-orbit coupling of Pt compared to (Co/Ni) combined to peculiar specular electronic reflections. \textit{Via} advanced simulations methods for the description of the spin-current profiles based on the spin-dependent Boltzmann formalism, we extracted the spin-Hall angle (SHA) of Pt and (Co/Ni) of opposite sign. The extracted SHA for Pt, +20\%, is opposite to the one of (Co/Ni), giving rise to an effective AHE inversion for thin (Co/Ni) multilayers (with the number of repetition layers $N<17$). The spin Hall angle in Pt is found to be larger than the one previously measured by complementary spin-pumping inverse spin-Hall effect experiments in a geometry of current perpendicular to plane. Whereas magnetic proximity effects cannot explain the effect, spin-current leakage and spin-orbit assisted electron scattering at Pt/(Co,Ni) interfaces fit the experiments. We also extract the main relevant electronic transport parameters governing the overall effects in current-in-plane (CIP) currents and demonstrate, in particular that the specularity/non-specularity in the electronic diffusion processes play an essential role to explain the results observed.
\end{abstract}

\pacs{72.25.Ba, 72.25.Mk, 75.70.Cn, 75.75.-}

\maketitle

\section{Introduction}

In recent years, the field of \textit{spinorbitronics} has emerged as a new route for spin-currents able to generate spin torques and excite small magnetic elements~\cite{liu2011a,liu2012a,liu2012b,pai2012,ohno2017}, to move domain walls~\cite{miron2011,boulle2013,emori2013,parkinano2015,rojas2016,baumgartner2017,Fukami2010}, to promote chiral Dzyaloshinskii-Moriya interactions (DMI)~\cite{boulle2013,chauleau2017} or to generate THz waves probed by the time domain spectroscopy (THZ-TDS) methods~\cite{munzenberg}. This is made possible \textit{via} the so-called \textit{intrinsic} spin Hall effect (SHE) and reciprocal inverse spin Hall effect (ISHE)~\cite{nagaosa2010,sinova2015,hellman2017} provided by heavy metals, \textit{e.~g.} Pt~\cite{hoffmann2013,jaffres2014,parkin2015,cornell2015,sagasta2016}, Ta~\cite{liu2012a,liu2012b} and W~\cite{hao2015} or \textit{via} the \textit{extrinsic} SHE of diluted metallic alloys investigated for their experimental properties~\cite{niimi2011,niimi2012,laczkowski2014,laczkowski2015,back2016,laczkowski2017,zhu2018,zhu2019,ralph2020} and from a theoretical point of view~\cite{fert1}. SHE borrows its concept from the well-established principles of the anomalous Hall effect (AHE)~\cite{nagaosa2010,sinova2015} whereby the relativistic spin-orbit interactions (SOI) may promote an asymmetric deflection of the electron flow depending on their spin. Early studies of AHE mostly deals with bulk ferromagnetic (FM) metals~\cite{karplus1954} and their alloys~\cite{smit1,smit2,Berger1970,fert1972,fert1981}.

With the fast development of spinorbitronics, AHE has started to be largely investigated in ultrathin multilayers such as Co/X from the beginning 90's with X=Au~\cite{Vavra1990}, Pd~\cite{Kim1993,Shaya2007,Rosenblatt2010,Guo2012,Kou2012,Keskin2013}, Pt~\cite{Canedy2000,Canedy2000b}, or more recently  Co/Ni multilayers~\cite{zhang2014} grown for their perpendicular magnetic anisotropy (PMA) properties~\cite{Arora2017} often required for magnetoelectronics devices. More recent experimental studies have exhibited the strong impact of the interfacial SOI on the injected spin-current at heavy metal/ferromagnetic metal interfaces thus promoting the necessary spin-orbit torques (SOT) for magnetic commutation~\cite{ralph2019}. This reveals the need to use accurate analyses and anatomy of spin-currents incorporating the spin-orbit degree of freedom.

To these ends, AHE and SHE involving 3\textit{d} transition metals (Fe, Co, Ni) or \textit{5d} noble metals are presently the basis of numerous fundamental investigations dealing with an intrinsic mechanism originating from the Berry phase~\cite{karplus1954,tanaka2008,guo2008}, skew-scattering diffusion processes~\cite{smit1,smit2,fert1981,mertig1,mertig2,casanova2018} and extrinsic side-jump~\cite{Berger1970,fert1972} phenomena. Moreover, very recent works dealt with the specific role of the AHE of CoNi~\cite{yuasa2020} and of NiFe~\cite{han2020} for a magnetization controlled spin-torques~\cite{amin2019,davidson2020}, the role of the electronic surface scattering on the properties of spin-current for AHE \textit{e.g.} at the PtO$_x$/Co interface~\cite{ando2019}; and the possible implication of the roughness on AHE for different related TM interfaces~\cite{manchon2020}. Examples of the relevance of such interface contribution are played by \textit{i}) the magnetic proximity effects in  Pt at the scale of few atomic planes~\cite{mokrousov2015,bailey2016,crowell2018,kelly2020}, as well as \textit{ii}) a possible spin-current depolarization or spin-memory loss (SML) at \textit{3d-5d} interface induced by local SOI as suggested and discussed recently~\cite{jaffres2014,nguyen2013,boone2015,nist2018,nist2018a,ralph2019,kelly2020}.
Regarding the issue of spin-current depolarization and its magnitude, another recent matter of debate is the typical value of the spin Hall angle (SHA) of \textit{5d} heavy metals such as Pt including both disorder~\cite{tanaka2007} and SML at interfaces~\cite{Cornell2016}.

Nonetheless, much less works have dealt with multilayered systems wherein interfaces may bring new insights in the spin-orbit assisted electronic scattering and diffusions. Like AHE, the spin-dependent SHE properties are scaled by the off-diagonal spin-dependent conductivity tensor $\sigma_{xy}^{s}$ involving either intrinsic ($\sigma_{xy}^{int,s}$)~\cite{chadova} or extrinsic skew-scattering ($\theta^{sk,s} \sigma_{xx}^{s}$) and extrinsic side-jump ($\sigma_{xy}^{sj,s}$) contributions~\cite{sinova2015}) with the result that:

\begin{eqnarray}
\sigma_{xy}^s=\sigma_{xy}^{int,s}+\theta^{sk,s} \sigma_{xx}^{s}+\sigma_{xy}^{sj,s}
\end{eqnarray}
where $s$ is the $\uparrow,\downarrow$ spin index, and where the subscripts $(sk),(sj),(int)$ denote respectively the \textit{extrinsic skew scattering, extrinsic side-jump} and \textit{intrinsic} contributive terms~\cite{chadova2}. In that frame, the off-diagonal conductivity term responsible for AHE that is the \textit{charge} contribution to AHE, $\sigma_{xy}$, is seen to be the sum of the two spin band channels, $\sigma_{xy}=\sigma_{xy}^\uparrow+\sigma_{xy}^\downarrow$, making the link between AHE and SHE we are searching for. AHE and SHE in \textit{3d} ferromagnetic materials, as in Co~\cite{kotzel2005,Hou2012} and Ni~\cite{lavine1961,Ye2012}, are mainly expected to possess an \textit{intrinsic} origin with opposite sign of $\sigma_{xy}^{int.}$ as experimentally calculated and determined~\cite{Wang2007,mokrousov2009,turek2012,weishenberg2011}. Nevertheless, the occurrence of exchange split Fermi surfaces in ferromagnets makes generally difficult to extract a clear relationship between AHE and SHE. This particular feature has been recently debated in a coupled or recent papers~\cite{mertig1,mertig2,casanova2018}.

\vspace{0.1in}

In the present work, we focus on the properties of AHE and on the control of the spin-current in \textit{standard} perpendicular magnetic configuration with the magnetization standing along the $z$ direction normal to the layers. Extensions could be made in the future to the case of the non-conventional situation of arbitrary magnetization direction within the ferromagnet~\cite{amin2019,davidson2020}. We present unconventional results and refined analyses of non-local AHE in a series of Pt/(Co0.2nm /Ni0.6nm)$_{N}$ and Au:W/(Co0.2nm/Ni0.6nm)$_{N}$ multilayers (MLs) involving different numbers of (Co/Ni) sequences and corresponding interfaces. (Co/Ni) is known to possess a specific interface anisotropy exhibiting PMA~\cite{Fukami2010, Mizukami2011,Arora2017}, and moreover involving Dzyaloshinskii-Moriya (DMI) interactions~\cite{parkin2014,chauleau2017}. By taking advantage of the relatively small AHE of (Co0.2/Ni0.6)$_N$ MLs and when $N<17$, we demonstrate, in some specific situation, a sign change of the AHE in a series of MLs structures grown onto the thin Pt buffer highlighted an opposite effective spin-orbit sign for Pt compared to Co/Ni. We used an advanced Boltzmann analysis algorithm for the necessary determination of the spin-currents with adequate boundary conditions at interfaces and based on the extension of the Fuchs-Sondheimer approach~\cite{fuchssondheimer} and like recently highlighted and exploited for (Co/Ni) multilayers~\cite{ohno2020}. It is based on the combination of electron scattering in multilayers~\cite{camley1989,ohno2020} and SOI-assisted spin and charge deflection inside layers. The different contributions have been carefully addressed involving off-diagonal spin-flip terms in the diffusive potentials. We thus demonstrate that the AHE sign inversion originates from the non-local properties of the spin-conductivities~\cite{Cornell2016,zhu2018} and the opposite sign of the effective SOI strength compared to the bulk (Co/Ni) MLs more than induced magnetization in Pt (magnetic proximity effects or MPE)~\cite{mokrousov2015,bailey2016,crowell2018}. Physical mechanisms of non-local AHE effect in those systems relies thus on the combination of the SOI-dependent scattering of a polarized current generated in (Co/Ni), specific electronic specularity reflections at interfaces~\cite{butler1995,stewart2003,chen17} and subsequent ISHE process in bulk adjacent heavy metal (Pt, Au:W). This also reveals a characteristic large positive SHA (+20\%) for Pt, at Co/Pt interface much stronger~\cite{parkin2015,Cornell2016,nist2018} than determined previously in combined spin pumping-ISHE experiments~\cite{jaffres2014} and that we may assign to an anisotropy in the scattering time~\cite{tanaka2007,tanaka2008}. By choosing consistent physical parameters, we find an excellent agreement with the experimental trends.

\vspace{0.1in}

\textcolor{black}{We have divided our paper into four different sections. The section \textit{II} is devoted the discussion of the sample preparation and experimental results dealing with Pt and Au:W buffer based multilayer (Co/Ni) samples. The section \textit{III} will be devoted to the description of the main modeling features governing the AHE and SHE phenomena possibly incorporating \textit{intrinsic} and \textit{extrinsic} phenomena. We present detailed calculation methods we used involving the spin-current profiles to model those effects in MLs before giving an accurate analysis of data emphasizing on the importance of the specularity at interfaces in section \textit{IV}. We give the main trends and conclusions in section \textit{V}.}

\section{Samples preparation and AHE experiments}

Samples are deposited on thermally oxidized Si wafers of two types, \textit{respectively} I and  II, at room temperature using magnetron sputtering. Type I and type II substrates differs by their level of roughness in the sub-nanometer scale like discussed just below. Samples are made of a 6-nm-thick heavy metal layer, Au:W alloys or Pt, covered by a magnetic multilayers composed of $N$ repetitions of (Co~0.2\,nm/Ni~0.6\,nm) bilayers with perpendicular magnetic anisotropy properties. Such (Co/Ni) stack is used to keep a large PMA, about constant at least up to $N\cong40-70$ in the present case in order to preserve a low roughness at the surface or even more ~\cite{Arora2017}. The resistivity $\rho$ of Au:W thin films varies from 80 to 130$\mu\Omega\cdot$cm depending on the W content in the alloy and we denote Au:W$_\rho$ the Au:W alloy of resistivity $\rho$ in $\mu \Omega\cdot$cm. The Pt bulk resistivity equals $\rho_{Pt}=17\mu \Omega\cdot$cm. All the data and parameters are gathered in Table I including those of Co, Ni and Al materials within the stack. Devices are then patterned into Hall cross bars of different widths ranging from 3 to 6~$\mu m$ and of 600~$\mu m$ length by optical UV lithography and Ar ion etching process. The samples are finally covered with a 5-nm-thick Al layer, hereafter oxidized onto 2~nm from the surface, to prevent oxidation of the (Co/Ni) stack. From previous work~\cite{laczkowski2017}, we can infer that Pt and Au:W buffer layers possess the same or opposite spin-Hall angle (SHA) depending on the W content in Au:W and depending thus of its resistivity $\rho_{Au:W}$. According to our convention, the SHA is counted positive, that of the same sign than that of Pt, for an Au:W alloy resistivity ($\rho$) typically less than $110~\mu \Omega\cdot$cm and, negative (same sign than Ni or Co0.2/Ni0.6) for $\rho> 120 \mu\Omega\cdot$cm. For type I \textcolor{black}{Si} substrates, Atomic Force microscopy (AFM) measurements allowed to determine an overall roughness of $0.3$~nm RMS at 1~$\mu m$ scale for (Co/Ni)$_5$ and 1.3~nm RMS for (Co/Ni)$_{20}$ samples. In contrast, for type II \textcolor{black}{Si} substrates, RMS is about 0.3~nm RMS for (Co/Ni)$_5$, 0.36~nm RMS for (Co/Ni)$_{20}$, 0.7~nm RMS for (Co/Ni)$_{70}$ and 1.1~nm RMS for (Co/Ni)$_{100}$. Then one must note that the RMS between substrates of type I and II are \textcolor{black}{not too different} for $N<5$, and then starts to strongly differ for $N>5$ up to $N=20-100$. However, samples all keep their strong PMA properties thus demonstrating the high quality of the nanometer scale interfaces, in particular free of large chemical intermixing.

\begin{figure}[tbp]
\includegraphics[width=12cm]{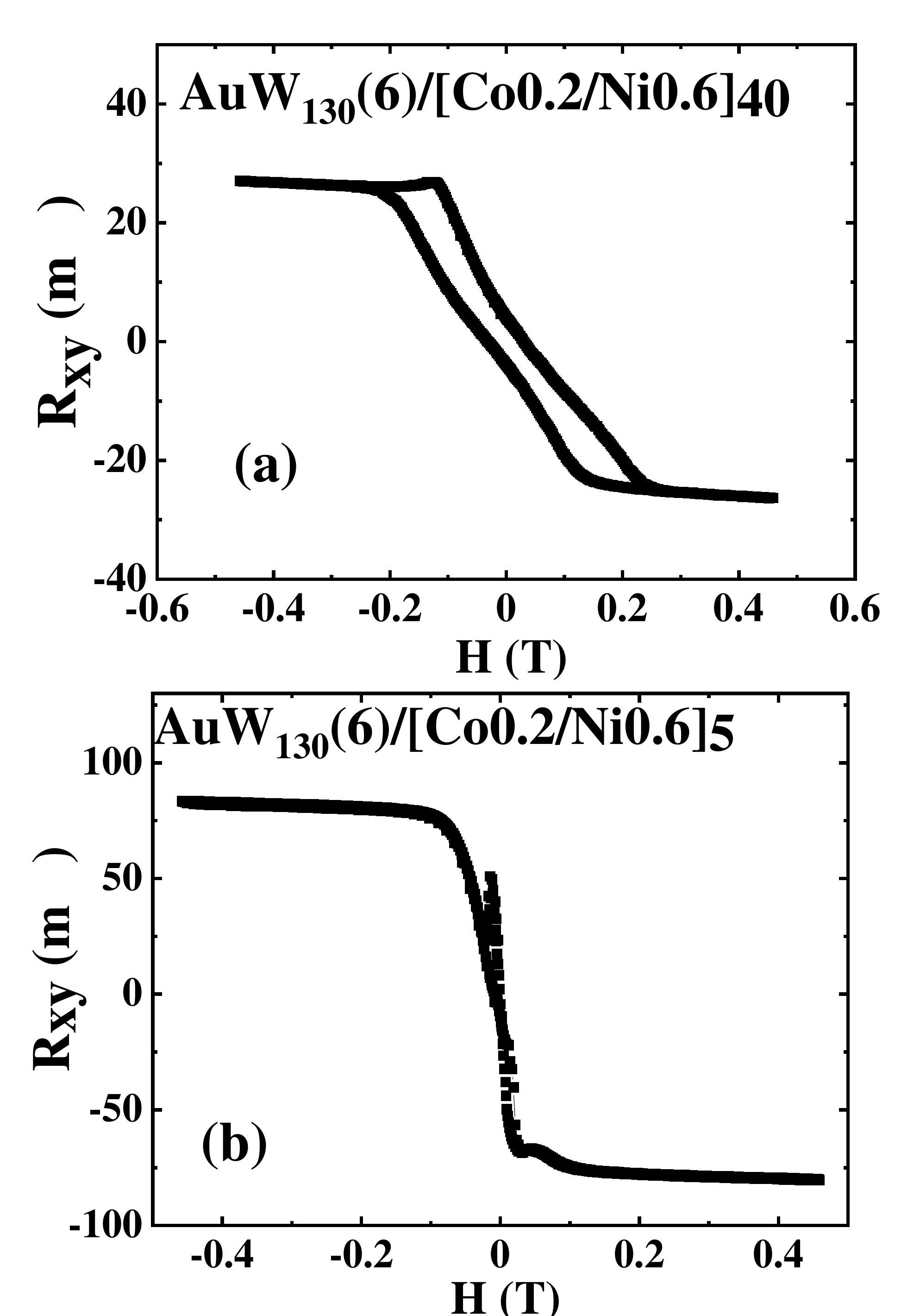}
\caption{Anomalous Hall effect (AHE) measurements acquired at room temperature (RT) on a)  Au:W$_{130}$6nm/(Co0.2nm/Ni0.6nm)$_{40}$ and b) Au:W$_{130}$6nm/(Co0.2nm/Ni0.6nm)$_{5}$ ML samples showing convention negative AHE sign for the (Co0.2/Ni0.6) multilayers. Samples are of type II and characterized by a large Au:W resistivity ($\rho\simeq 130 \mu \Omega$~cm).}
\label{fig:1}
\end{figure}

\textcolor{black}{We discuss below the main AHE results obtained on the different set of samples where a small ordinary Hall effect (OHE) has been subtracted in each case.  OHE, preferentially revealed at low temperature (LT), will be briefly discussed hereafter and displayed in the supplemental information~\cite{supplemental}.}

Fig.~\ref{fig:1} displays the AHE results \textcolor{black}{obtained at room temperature (RT) on} type II samples (very flat) made of Au:W$_{130}$ 6~nm buffer of a large resistivity, $\rho\simeq ~130 \mu \Omega$~cm, and characterized by a SHE sign opposite to the one of Pt~\cite{laczkowski2017}. Fig.~\ref{fig:1}a shows the transverse resistance measurements, \textit{i.e.} the AHE of the $Au:W_{130}$ 6/(Co0.2/Ni0.6)$_{40}$ device with a large period repetition $N$=40 and corresponding to an overall (Co,Ni) thickness of 32~nm that we can consider as \textit{bulk} (Co/Ni). AHE amplitude of $\Delta R_{AHE}$=-28~m$\Omega$ for $N=40$ gives a negative signature for the AHE of Ni following our convention. This remains also true for (Co/Ni) MLs when Ni, of a large intrinsic AHE, is thicker than Co and when $N$ is sufficiently large for (Co/Ni)$_N$ to dominate the conduction process. From Fig.~\ref{fig:1}b, the same conclusions holds for the samples grown on the 6~nm Au:W$_{130}$ buffer layers for $N$=5 associated, \textcolor{black}{in proportion}, to a larger current shunt in the Au:W$_{130}$. The AHE amplitude is of the same sign (negative) and characterized by a larger amplitude $\Delta R_{AHE} $=-80~m$\Omega$ thus demonstrating the particular role of the buffer on the spin-current properties. This observation will be ascribed, \textit{via advanced calculation methods} (section IV), to a larger SHE angle of Au:W$_{130}$ compared to (Co/Ni) and of the same sign, allowing thus a more efficient spin-charge conversion (AHE signal) of the spin-polarized current generated from the (Co/Ni) ferromagnetic layer.

\begin{figure}[tbp]
\includegraphics[width=15cm]{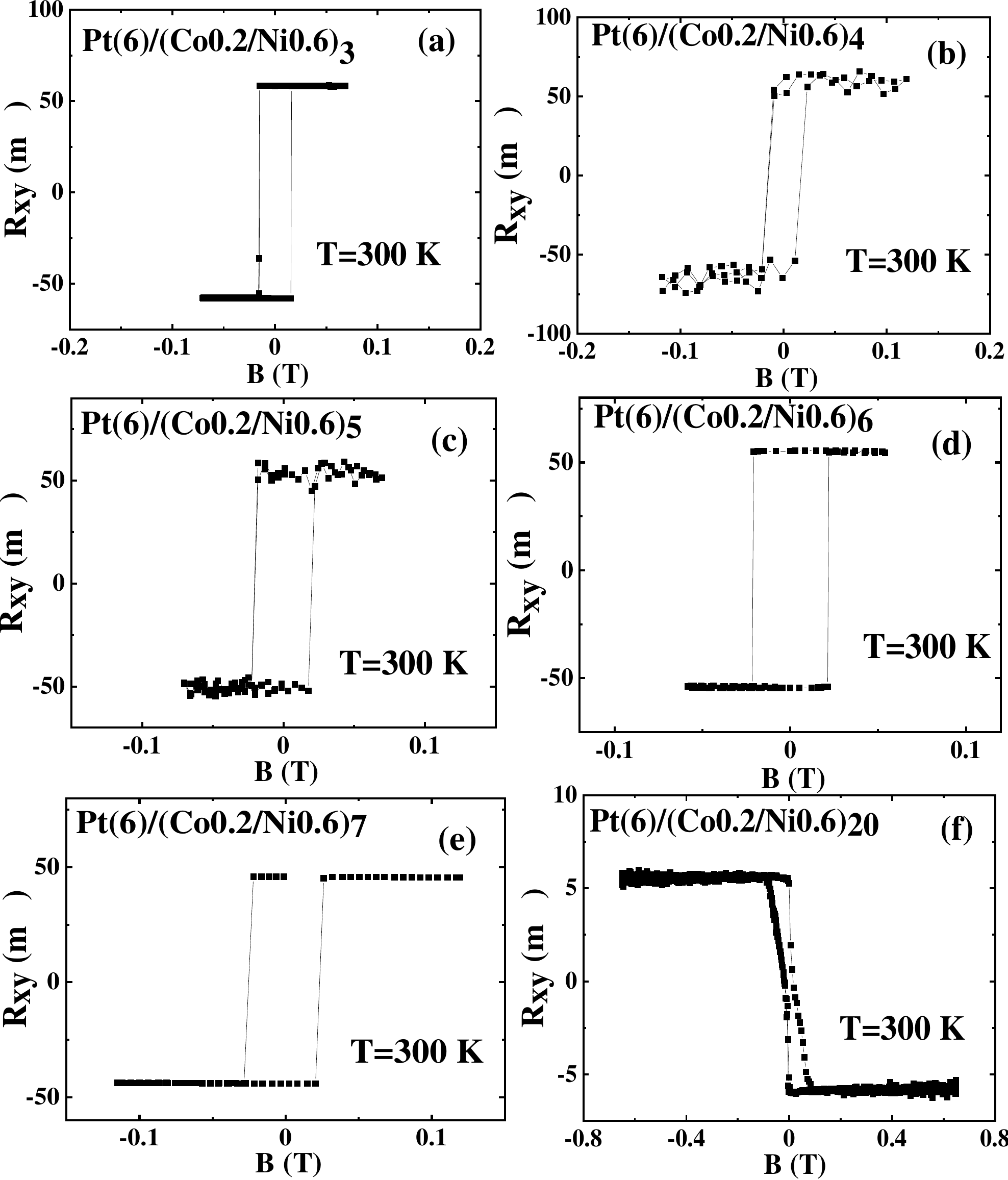}
\caption{Anomalous Hall effect (AHE) measurements acquired at room temperature (RT) on the type II Pt6nm/(Co0.2nm/Ni0.6nm)$_N$ series for N=3 (a), 4 (b), 5 (c), 6 (d), 7 (e) and 20 (f). For N=3-7, samples display a positive AHE whereas for N=20 the AHE is negative.}
\label{fig:2}
\end{figure}

\begin{figure}[tbp]
\includegraphics[width=1\linewidth]{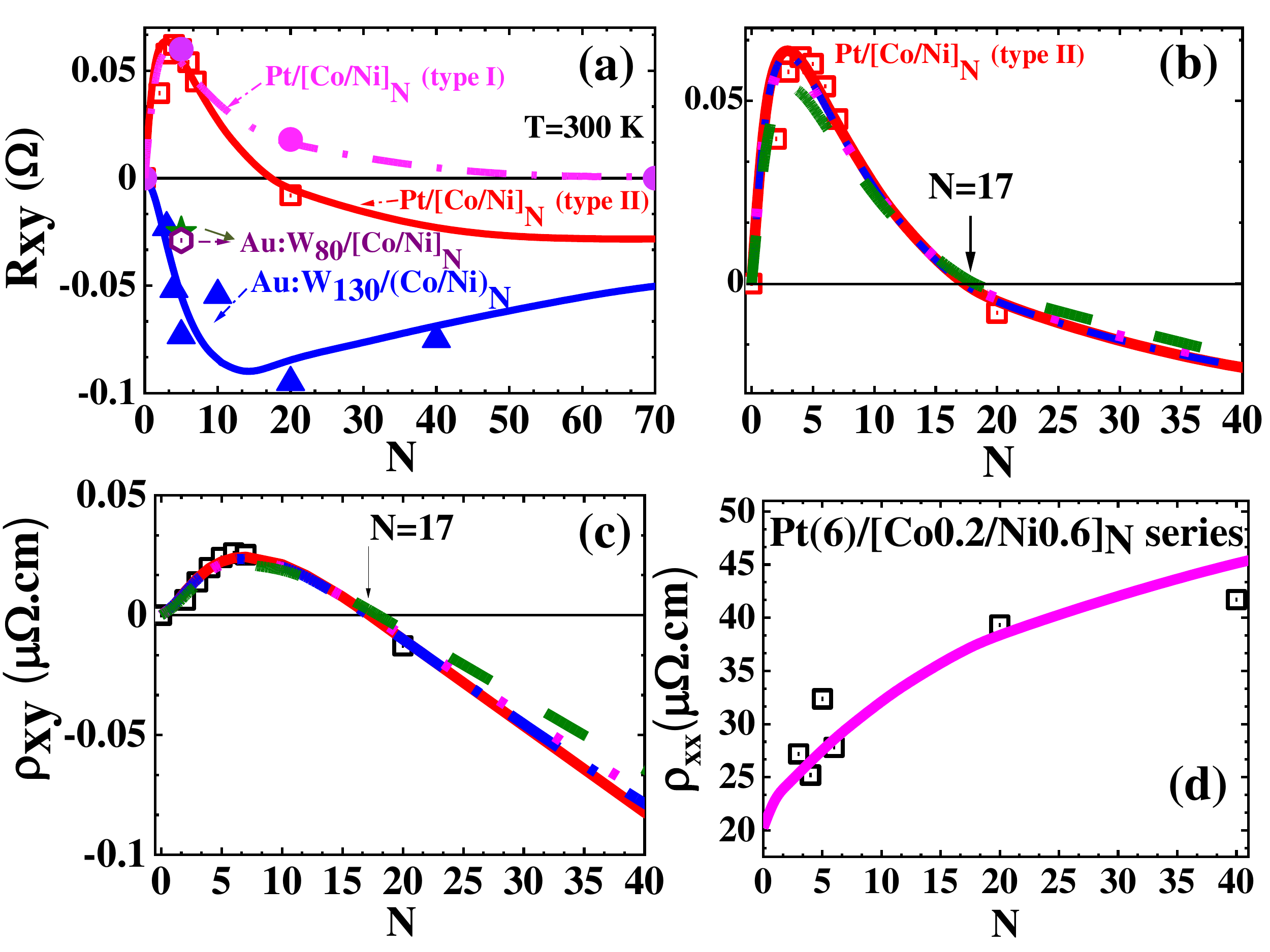}
\caption{a) AHE resistance and resistivity \textit{vs.} the number of sequence $N$ of (Co/Ni) for a) the Pt type I (orange dots) and type II series (red squares). The AHE resistance for the Au:W$_{130}$ series (black triangle) and Au:W$_{80}$ (purple point) compared to Pt are also reported. b) Different fits for type II Pt series are represented in lines corresponding to intrinsic AHE in (Co/Ni) (red line) or different models for extrinsic SHE (black, purple, green) -refer to section IV for the discussion-. c) corresponding AHE resistivities $\rho_{xy}$ \textit{vs.} $N$ for type II Pt (black squares). (d) longitudinal resistivities of Pt type II samples \textit{vs.} $N$ compared to the resulting fit (red line). The ensemble of parameters used for the fit of type II Pt are gathered in Table I.}
\label{fig:3}
\end{figure}

On Figs.~\ref{fig:2}(a-f), we report on the AHE \textcolor{black}{acquired at RT} provided with the same type II samples series involving a Pt buffer. They \textit{black}{are} made of Pt 6~nm buffer with the number $N$ of (Co/Ni) repetition layers within the $N=3-20$ range and fabricated within the same batch. Figs.~\ref{fig:2} from (a) to (f) refers respectively to $N=3~(a), 4~(b), 5~(c), 5~(d), 7~(e)$ and $20~(f)$. It is now quite remarkable that AHE for type II Pt-based buffer becomes positive for $N=3-7$, that is of the opposite same sign compared to the Au:W type I reference samples series described above.  In that sense, such sign inversion constitutes a major AHE properties identified in the present investigations. The AHE signal, \textcolor{black}{now positive}, first increases in amplitude from $\Delta R_{AHE}^{(2)}$=+40~m$\Omega$ for $N=3$ to $\Delta R_{AHE}^{(5)}$=+60~m$\Omega$ for $N=5$ before decreasing for $N=6, 7$ to $\Delta R_{AHE}^{(6)}$=+54~m$\Omega$ and $\Delta R_{AHE}^{(7)}$+=45~m$\Omega$ respectively. Moreover for $N$=20, the AHE becomes negative that is acquires the same sign than the Au:W$_{130}$ reference sample with an amplitude equal to $\Delta R_{AHE}^{(20)}$=-8~m$\Omega$. This signature of a sign change for the Pt series has to be assigned to an apparent opposite sign of the spin-orbit interactions (SOI) in Pt, at least for $N=3, 7$ compared to (Co/Ni) and Au:W samples. Moreover, the existence of a maximum in the AHE size for $\Delta R_{AHE}^{(5)}$ may be explained by a cross-over between \textit{i)} the positive contribution of the SOI in Pt with respect to (Co/Ni) and \textit{ii)} the needs of a spin-polarized current provided by the (Co/Ni) ferromagnet MLs as explained in section IV. \textcolor{black}{At this stage, two additional remarks have to be appended: \textit{i)} the sign inversion of AHE for the $N=3-7$ Pt series is independent of the temperature effects (not shown), as well as the sign inversion of AHE in Pt observed between $N$=3-7 and $N$=20}~\cite{supplemental}. This makes our results and conclusion independent of thermal effects~\cite{casanova2018}, and \textit{ii}) identical AHE experiments led with type I (very flat) Pt samples exhibit the same trends than type II with typical AHE sign inversion for \textcolor{black}{$N=3-7$}. The whole experimental results concerning AHE for the three samples series (Pt and Au:W type I and Pt type II series) are gathered on Fig.~\ref{fig:3}.

\textcolor{black}{A small OHE effect can be revealed at LT (10~K) on the different samples and all shows the same NHE slope in sign with about the same amplitude corresponding to a Hall coefficient in the range $-1,1.5\times 10^{-12}$~V.cm/(A.G) in the exact range of what is expected from Co or Ni at RT~\cite{volkenshtein60}. The same sign of NHE for (Co/Ni) accompanied by a sign inversion for AHE represents another clear indication of the AHE reversal by proximity effects and SOI sign inversion.}

Fig.~\ref{fig:3} displays both AHE resistances, effective AHE resistivities, as well as longitudinal resistivities \textcolor{black}{for the whole series} \textit{vs.} the number sequences $N$ varying from 3 to 70. In those experiments, the structure of (Co0.2/Ni0.6) bilayers is fixed depending only on $N$. Fig.~\ref{fig:3}a highlights the typical crossover from positive to negative experimental values acquired at RT, for the AHE resistance $R_{xy}$ in the case of Pt type II. The crossover between positive to negative AHE for type II is obtained for $N\simeq 17$ corresponding to a total (Co/Ni) thickness of about 15~nm. This point indicates an exact compensation of the AHE current contribution provided by Pt (positive AHE) and (Co/Ni) (negative AHE) layers. On the other hand, one can observe on the same Fig.~\ref{fig:3}a, that the $R_{xy}$ of type I Pt series follows the same features than type II Pt series (AHE inversion) with about equal values, before decreasing to zero and crossing to negative values for larger $N>70$ (orange points and curve \textit{black}{corresponding to N=70}). This demonstrates the strong impact of the roughness on the AHE amplitude and sign in thin multilayered samples.

Fig.~\ref{fig:3}c gives the same plot than Fig.~\ref{fig:3}a in the scale of resistivity $\rho_{xy}$ for type II series whereas Fig.~\ref{fig:3}d displays the typical expected increase of the resistivity in the Pt type II series from the one of pure Pt ($17~\mu \Omega.$cm) to the one of (Co/Ni) (about $50~\mu \Omega.$cm) at RT when $N$ increases to saturate at the bulk value of (Co/Ni). On the other hand, Fig.~\ref{fig:3}b moreover compares the AHE values obtained with both Pt and different Au:W series with Au:W$_{130}$ (black triangle) and Au:W$_{80}$ (purple point) characterized by a lower resistivity ($80~\mu \Omega.$cm) and opposite spin-Hall angle (SHA) equal to $\theta_{Au:W_{80}}$=+0.15~\cite{laczkowski2017}. The Au:W$_{80}$ experimental point lies between the corresponding Pt and Au:W$_{130}$ samples because of its intermediate SHA value between the ones of the two materials. One of our main conclusions is that, for thin ferromagnetic stacks ($N$ small), AHE may be strongly dependent on the heavy-metal buffer grown for PMA properties as well as the surface roughness and possibly, the specularity at the different interfaces of the electronic waves.

\section{Main physical issues of AHE and SHE in transition metals}

In that section, we discuss the main fundamentals prerequisites for the description of \textit{i)} AHE of 3\textit{d} transition metals on the basis of spin-currents and spin-channel dependent inverse SHE (ISHE) and subsequently \textit{ii)} ISHE in 5-\textit{d} transition heavy metals like Pt or Ta. Large details can be found in previous reference works~\cite{karplus1954,smit1,smit2,Berger1970,fert1981,tanaka2007,tanaka2008,guo2008,weishenberg2011}, review articles~\cite{sinova2015,nagaosa2010} and also in more recent contributive works~\cite{mertig1,mertig2,casanova2018,chadova,wang2019}.

\subsection{Linear Response theory for SHE/AHE}

What are the main mechanisms responsible for spin current generation in bulk \textit{3d} ferromagnets and their possible extension to spin-orbit assisted interface scattering? AHE, discovered in 1880 by Edwin Hall, precedes many of the spin-orbit effects being studied today. It  describes a large magnetization-dependent Hall effect in a ferromagnetic conductor and may be decomposed into extrinsic mechanisms originating from electronic quantum diffusions and intrinsic mechanisms originating from quantum dynamics within the host material. The description of the spin-current properties and spin-dependent conduction or conductivity may be, at a first stage, handled \textit{via} a $2\times 2$ band model within the linear response theory adapted from Kubo's approach. The systems accounts for an Hamiltonian of the type:

\begin{equation}
\hat{H}=\frac{\hbar^2 \hat{\mathbf{p}}^2}{2m^*}+\hat{V}(\mathbf{r})+\frac{\lambda_{SO}}{\hbar}~\left(\hat{\nabla} \hat{V}(\mathbf{r}) \times  \hat{\mathbf{p}}\right)\cdot\hat{\sigma}
\end{equation}
written in a 2-band spinor form and where $\mathbf{k}$ is the electronic wave-vector, $\hat{V}(\mathbf{r})$ the local spherical energy potential experimented by the \textit{sp-d} electron ($\hat{V}$ stands then for the potential energy $eV$ for the sake of simplicity) and $\lambda_{SO}$ the spin-orbit strength in unit of a (spin-orbit) cross-section or area. Inside $V(\mathbf{r})$ is also included the external electron potential responsible for the electron drift induced by the external electric field $\mathbf{E}$. Moreover, inside $\lambda_{SO}~\hat{\sigma}\cdot\left(\hat{p}\times \nabla eV(\mathbf{r})\right)$, it may be included the Rashba Hamiltonian arising from the action of $\mathbf{E}$ at interfaces on the electronic band structure. We note respectively $|\mathbf{k},n,s>$ and $|\mathbf{k}^\prime,m,s^\prime>$ (with $\mathbf{p}=\hbar \mathbf{k}$), the respective ingoing and outgoing electronic states diffused away from a scattering center and acting as a local spin-orbit perturbation $\delta \hat{V}(\mathbf{r})$ from the host. $n$ and $m$ are the band index whereas $s=\uparrow,\downarrow$ and we remind that $s^\prime=\uparrow,\downarrow$ is the spin index.

\vspace{0.1in}

The anomalous Hall conductivity (AHC) for transition metals ferromagnets may be derived from the sum of the spin-selected conductivites whereas the spin-Hall conductivity (SHC) derives from its difference. The selected spin-conductivity $\sigma_{xy}^s$ then may be written.

\begin{equation}
\sigma_{xy}^{s}=\frac{\sigma_{xy}^{AHE}\pm \sigma_{xy}^{SHE}}{2}
\end{equation}
with the $+$ ($-$) sign corresponding to the spin $\uparrow$ (spin $\downarrow$).  One must be care with the particular geometry configuration of a magnetization aligned along the normal direction to the layer ($z$). When the magnetization is rotated away from the $z$ direction, other types of spin-current, expressed in a vectorial form, may appear as the spin anomalous Hall effect (SAHE)~\cite{amin2019,davidson2020}.

Within the linear response theory, the Kubo formalism relates both the conductivity and the spin-conductivity to the equilibrium current-current and current spin-current correlation functions. This techniques provides a fully quantum mechanical formally exact expression for the conductivity within the linear response theory~\cite{chadova}. We emphasize the key issues in investigating the AHE/SHE properties within this formalism in order to adapt it for the case of multilayered systems. For the purpose of studying the AHE/SHE, it is worth to formulate the current-current and the spin current-current correlation within the Kubo formula in the form of the so-called Kubo-Bastin formula. If one denotes $\hat{v}_\alpha=\frac{p_\alpha}{m^*}$ the velocity along $\alpha$ and $\{\hat{v}_\beta \sigma^\gamma\}=\frac{\hat{v}_\beta \sigma^\gamma+\sigma^\gamma \hat{v}_\beta}{2}$  the spin-current flowing along the $\beta$ direction with spin directed along the $\gamma$ axis in a symmetrized form ($\sigma^0=\hat{I}$ the unit matrix for the charge conductivity), the Kubo-Bastin formula reads:

\begin{widetext}
\begin{eqnarray*}
\sigma_{\alpha \beta}^\gamma=-\frac{e^2\hbar}{2\pi\mathcal{V}}\int f(\epsilon)Tr <\hat{v}_\alpha \frac{\partial G^R(\epsilon)}{\partial \epsilon}\{\hat{v}_\beta \sigma^\gamma\} (G^R(\epsilon)-G^A(\epsilon))-\hat{v}_\alpha(G^R(\epsilon)-G^A(\epsilon)) \{\hat{v}_\beta \sigma^\gamma\}\frac{\partial G^A(\epsilon_F)}{\partial \epsilon}> d \epsilon
\label{kubobastin}
\end{eqnarray*}
\end{widetext}
where $G^{R,A}=\frac{1}{\epsilon-\hat{H}\pm i\Gamma}$ is the respective retarded (R) and advanced (A) Green's function with $\epsilon$ is the energy, $\Gamma$ the energy broadening, $f(\epsilon)$ is the Fermi occupation function and $Tr$ is the trace over $\mathbf{k}$ and band index $n$. By integration the previous integral by parts, the spin-conductivity breaks up into $\sigma_{\alpha\beta}^\gamma=\sigma_{\alpha\beta}^{\gamma,I}+\sigma_{\alpha\beta}^{\gamma,II}$ where $\sigma_{\alpha\beta}^{\gamma,I}$ and $\sigma_{\alpha\beta}^{\gamma,II}$ stand respectively for the Fermi surface (spin-)conductivity term and $\sigma_{\alpha\beta}^{\gamma,II}$ for the Fermi sea (spin-conductivity) term according to:

\begin{widetext}
\begin{eqnarray*}
\sigma_{\alpha\beta}^{\gamma,I}=\frac{e^2\hbar}{2\pi\mathcal{V}}Tr<\hat{v}_\alpha G^R(\epsilon_F)\{\hat{v}_\beta \sigma^\gamma\}G^A(\epsilon_F)>-\frac{1}{2}<R\rightarrow A+A\rightarrow R>
\label{fermisurface}
\end{eqnarray*}
\end{widetext}
and:

\begin{widetext}
\begin{eqnarray*}
\sigma_{\alpha\beta}^{\gamma,II}=\frac{e^2\hbar}{4\pi\mathcal{V}}\int d\epsilon f(\epsilon) Tr<\hat{v}_\alpha G^R(\epsilon)\{\hat{v}_\beta \sigma^\gamma\}  \frac{\partial G^R(\epsilon)}{\partial \epsilon}-\hat{v}_\alpha \frac{\partial G^R(\epsilon)}{\partial \epsilon}\{\hat{v}_\beta \sigma^\gamma\}  G^R(\epsilon)+c.c>
\label{fermisea}
\end{eqnarray*}
\end{widetext}

\vspace{0.1in}

SHC and AHC originate from the same general formula representative of the linear response theory. The  difference lies in that the involves a symmetrization ($\{\}$) procedure because involving non-commutative Pauli matrices for the spin. In that sense, both quantities, SHE and AHE, are connected each-other.

We are going to describe the different contributions to the spin-current or spin-Hall conductivities (SHC) in terms of \textit{extrinsic} \textit{vs.} \textit{intrinsic} effect. Kontani \textit{et al.}~\cite{tanaka2007} and Tanaka \textit{et al.}~\cite{tanaka2008} conclude that the origin of SHC in transition metals, 3\textit{d} ferromagnet or 5\textit{d} heavy metals, depends on the degree of disorder and on the characteristic energy broadening ($\Gamma$). At small disorder (small $\Gamma$), SHC is intrinsic and originates mainly from the contribution Fermi sea term $\sigma_{xy}^{II}$ (interband contribution). On the other hand, as the disorder increases (large $\Gamma$), the Fermi Fermi surface term (intraband contribution) becomes dominant.

\subsubsection{intrinsic AHE/SHE.}
The intrinsic spin-conductivity mainly originates from the Fermi sea terms and it can be derived from the expansion of the Kubo-Bastin formula~\ref{kubobastin} for the host. It writes in the general case as:

\begin{eqnarray*}
\sigma_{\alpha\beta}^{\gamma}=2\frac{e^2\hbar}{\mathcal{V}}\Im m\sum_{kn,km}\left[f\left(\epsilon_n\right)-f\left(\epsilon_m\right)\right]\frac{<kn|\{\hat{\sigma}^{\gamma}\hat{v}_\alpha\}|km><km|\hat{v}_\beta|kn>}{(\epsilon_{nk}-\epsilon_{mk})^2}
\end{eqnarray*}
where $f$ is the Fermi occupation function, $|n>$ and $|m>$ are respectively \textit{occupied} and \textit{unoccupied} electronic state, $\epsilon_{km}-\epsilon_{kn}$ the corresponding excitation energy, $\hat{v}$ is the velocity operator, $\hat{\sigma}$ the Pauli matrix, $\alpha,\beta$ two indices for the space coordinates, $xx$ and $xy$ for the respective longitudinal and transverse spin Hall conductivity, and $\Im m$ the imaginary part. When $\gamma=z$ defines the spin direction along z, it leads \textit{in fine} to~\cite{guo2008}:

\vspace{0.1in}

\begin{eqnarray*}
\sigma_{xy}^{\gamma}=2\frac{e^2\hbar}{\mathcal{V}}\Im m\sum_{n}f\left(\epsilon_n\right)\Omega_{(n)}^{z}
\end{eqnarray*}
where $\Omega_{(n)}^{z}=\Im m\sum_{km\neq n}\frac{<kn|\{\hat{\sigma}^{z}\hat{v}_x\}|km><km|\hat{v}_y|kn>}{(\epsilon_{kn}-\epsilon_{km})^2}$ stands for the Berry curvature corresponding to band $n$.

\subsubsection{Extrinsic skew scattering AHE/SHE.}

We are now going to describe AHE/SHE in the limit of strong scattering. Smit~\cite{smit1,smit2} proposed the skew-scattering mechanism as the source of the AHE. Indeed, in the presence of spin-orbit interactions (SOI), the matrix element of the impurity scattering potential reads:

\begin{equation}
<\mathbf{k}^\prime s^\prime|\delta\hat{V}|\mathbf{k} s>=\tilde{V}_{\mathbf{k},\mathbf{k}^\prime}\left[\delta_{s,s^\prime}+\lambda_{so} <s^\prime|\hat{\sigma}|s>\cdot \left(\mathbf{k}^\prime \times \mathbf{k}\right)\right]
\end{equation}
for a spherical perturbation $\delta\hat{V}$. Microscopic detailed balance would require that the transition probability $W_{n \rightarrow m}$ between states $|k,n,s>$ and $|k^\prime,m,s>$ is identical, at a second order of perturbation, to that proceeding in the opposite direction $W^{(2)}_{m \rightarrow n}$. It holds for the Fermi's golden-rule approximation,

\begin{equation}
W_{k \rightarrow k^\prime}=\frac{2\pi}{\hbar}|<\mathbf{k}^\prime s^\prime|\delta\hat{V}|\mathbf{k} s>|^2  \delta\left(\epsilon_{\mathbf{k}n}-\epsilon_{\mathbf{k}^\prime}\right)
\end{equation}

At higher order, the transition rate $W_{\mathbf{k}\mathbf{k}^\prime}$ is given by the T-matrix element of the disorder potential $W_{\mathbf{k}\mathbf{k}^\prime}\simeq \frac{2\pi}{\hbar}|T_{\mathbf{k}\mathbf{k}^\prime}|^2 \delta\left(\epsilon_{\mathbf{k}^\prime}-\epsilon_{\mathbf{k}}\right)$. The scattering T matrix is more generally defined as $T_{\mathbf{k}\mathbf{q}^\prime}=<\mathbf{q}^\prime|\delta\hat{V}|\mathbf{k}>$ with $\mathbf{q}$ being the eigenstate of the full Hamiltonian involving the perturbation $\delta \hat{V}$. For a weak disorder, one can approximate the scattering state $|\mathbf{q}^\prime>$ by a truncated series in powers of $\delta V_{\mathbf{kk}^\prime}$ according to:

\begin{eqnarray}
|\mathbf{q}>=|\mathbf{k}>+\Sigma_{\mathbf{k}^{\prime\prime}}\frac{\delta V_{\mathbf{k}^{\prime\prime}\mathbf{k}}}{\epsilon_{\mathbf{k}}-\epsilon_{\mathbf{k}^{\prime\prime}}+i\eta}|\mathbf{k}^{\prime\prime}>
\end{eqnarray}

Using this expression in the above definition of the $T$ matrix, one can expand the scattering rate in powers of the disorder strength up to the order 2 (leading to no scattering asymmetry terms) or upt to the order three (leading to scattering asymmetry terms) giving \textit{in fine}:

\begin{equation}
W_{\mathbf{k}\mathbf{k}^\prime}^{(3)}=\frac{2\pi}{\hbar} \left(\Sigma_{\mathbf{k}^{\prime\prime}} \frac{<\delta \hat{V}_{\mathbf{k}^\prime\mathbf{k}^{\prime\prime}}\delta \hat{V}_{\mathbf{k}^{\prime\prime}\mathbf{k}}\delta\hat{V}_{\mathbf{k}^{\prime}\mathbf{k}}>}{\epsilon_\mathbf{k}-\epsilon_{\mathbf{k}^{\prime\prime}}+i\eta}+c.c \right)\delta(\epsilon_{\mathbf{k}}-\epsilon_{\mathbf{k}^\prime})
\end{equation}
which yields \textit{in fine}:

\begin{equation}
W_{\mathbf{k} \mathbf{k}^\prime}^{(3)}=-\frac{(2\pi)^2}{\hbar} \lambda_{SO} \mathcal{N}(\epsilon_F) \hat{V}^3 \left(\mathbf{k}^\prime\times \mathbf{k}\right)\cdot\hat{\mathbf{\sigma}} \delta(\epsilon_{\mathbf{k}}-\epsilon_{\mathbf{k}^\prime}) \delta_{\sigma \sigma\prime}
\end{equation}
for the asymmetry term we are searching for ($\Im$ is the imaginary part) and $\mathcal{N}(\epsilon_F)$ is the density of states at the Fermi level. In the calculations of the Hall conductivity, which involve the second Born approximation (third order in $V$), detailed balance already fails. In the case of \textit{p}-orbitals, skew scattering can be represented by an asymmetric part of the transition probability

\begin{equation}
W_{\mathbf{k}\mathbf{k}^\prime}^A=-\tau_A^{-1}(\mathbf{k}\times \mathbf{k}^\prime)\cdot \hat{\mathbf{m}}
\end{equation}
When the asymmetric scattering processes is included (called skew scattering), the scattering probability $W_{\mathbf{k} \mathbf{k}^\prime}$ is distinct from $W_{\mathbf{k} ^\prime \mathbf{k}}$. Physically, scattering of a carrier from an impurity introduces a momentum perpendicular to both the incident momentum $k$ and the magnetization \textbf{m}. This leads to a transverse current proportional to the longitudinal current driven by the electric field \textbf{E}.

\subsubsection{Extrinsic side jump AHE and SHE.}

The side-jump mechanism mainly originates from the Fermi surface term. The basic semi-classical argument is the following: when considering the scattering of a Gaussian wave packet from a spherical impurity with SOI (with $H_{SO}=\lambda_{SO} ~L_z ~S_z$, a wave packet with an incident wave vector $\mathbf{k}$ will experiment a displacement transverse to $\mathbf{k}$ and $\mathbf{E}$ 
This type of contribution was first noticed, but discarded by smit~\cite{smit2}and reintroduced afterwards by Berger who argued that it was the main contribution to the AHE. The derivation of the side-jump process can be made by the following semi-phenomenological arguments. The change in the velocity during the diffusion process by a central potential $\delta \hat{V}(r)$ simply writes:

\begin{equation}
\hat{\mathbf{v}}_i=\frac{\partial \hat{\mathbf{r}}_i}{\partial t}=\frac{1}{i\hbar}[\hat{\mathbf{r}}_i,\hat{H}]=\frac{\mathbf{\hat{p}}_i}{m^*}-\frac {\lambda_{SO}}{\hbar}\left(\nabla \delta \hat{V}(r)\times \hat{\sigma}\right)
\end{equation}
However, the dynamics of the electron is also described by:

\begin{equation}
\frac{\partial \hat{\mathbf{p}}_i}{\partial t}=\frac{1}{i\hbar}[\hat{\mathbf{p}}_i,\hat{H}]\simeq-\nabla \delta \hat{V}(r)
\end{equation}
ans we get the anomalous Hall velocity:

\begin{equation}
\hat{\mathbf{v}}_i=\frac{\hat{\mathbf{p}}_i}{m}+\frac {\lambda_{SO}}{\hbar}\left(\frac{\partial \hat{\mathbf {p}}}{\partial t}\times \hat{\sigma} \right)_i
\end{equation}
The time-integration of the last quantity gives out the lateral shift coordinate $\Delta r_i$ ($i$ represents here the transverse direction) experienced by a carrier during a scattering event:
\begin{equation}
\Delta \mathbf{r}_i=\frac{\lambda_{SO}}{\hbar}\left(\Delta \mathbf{p} \times \hat {\sigma}\right)_i
\end{equation}
and the transverse side-jump velocity $\mathbf{v}_{sj}$ results in:

\begin{equation}
\mathbf{v}_{sj}=\frac{\lambda_{SO}}{\hbar}\frac{\Delta \mathbf{p} \times \hat{\sigma}}{\tau_p}
\end{equation}
with $\tau_p$ is the characteristic momentum scattering time and $\Delta \mathbf{p}=e\mathbf{E}\tau_p$ the change of the particle impulsion after its isotropic diffusion. We get the side-jump current along the direction normal to both electric field $\mathbf{E}$ and spin-direction $\sigma$ according to:

\begin{equation}
\mathbf{j}^{sj}=\frac{ne^2\lambda_{SO}}{\hbar}\left(\mathbf{E} \times \hat{\sigma} \right)
\end{equation}
and where $n$ is the density of carriers. Note that the side-jump mechanism, like the intrinsic AHE/SHE is independent on the scattering time. A detailed expression of the anomalous Hall velocity in a multiband picture for TM have been derived by Levy~\cite{levy1988} using phase-shift analyses.

\subsection{Model for AHE and SHE for 3\textit{d} ferromagnets and 5\textit{d} heavy metals}

From the fundamental principle described above, we now turn on the description of spin Hall effects (SHE) and anomalous Hall effect in transition metals \textit{3d} ferromagnets. This connection is described by the off-diagonal terms of conductivity tensors $\sigma_{xy}$ for a spin direction aligned along $\hat{z}$: whereas SHE is described \textit{via} the spin-dependent ($\uparrow$, $\downarrow$ spins) conductivity tensors, the AHE considers the sum of those two spin-dependent contributions according to:

\begin{eqnarray}
\hat{\sigma}_{\alpha\beta}=\hat{\sigma}_{\alpha\beta}^\uparrow+\hat{\sigma}_{\alpha\beta}^\downarrow
=\left(
\begin{array}{cc}
\sigma_{xx}^{\uparrow}+\sigma_{xx}^{\downarrow} & \sigma_{xy}^{\uparrow}+\sigma_{xy}^{\downarrow}\\
-(\sigma_{xy}^{\uparrow}+\sigma_{xy}^{\downarrow}) & \sigma_{xx}^{\uparrow}+\sigma_{xx}^{\downarrow}
\end{array}
\right)
\end{eqnarray}
where $\sigma_{xx}^{\uparrow,\downarrow}$ represents the longitudinal conductance along the electric field direction ($x$) and $\sigma_{xy}^{\uparrow,\downarrow}$, the off-diagonal or transverse part responsible for both SHE and AHE phenomena with transverse particle flow along $y$. Accordingly, SHE are both described by their own angle related by:

\begin{eqnarray}
\theta_{SHE}^{s}=\frac{\sigma_{xy}^{s}}{\sigma_{xx}^{s}} \qquad;\qquad
\theta_{AHE}=\frac{\sigma_{xy}^{\uparrow} +\sigma_{xy}^{\downarrow}}{\sigma_{xx}^{\uparrow} +\sigma_{xx}^{\downarrow}}
\end{eqnarray}
Ferromagnetic materials like Co, Ni or (Co/Ni), $\theta^{\downarrow}$ differs from $\theta^{\uparrow}$ due to the lift in the energy spin-degeneracy, particularly at the Fermi level~\cite{zimmermann} whereas for non-magnetic materials, like Pt, one has $\theta^{\downarrow}=-\theta^{\uparrow}$ by simple symmetry arguments. The last expression for the AHE angle gives the correspondence between respective AHE and SHE mechanisms. The spin-dependent conductivity and resistivity tensors are inverse each others. One considers the respective off-diagonal components which mainly writes as a sum of three different physical contributions, \textit{extrinsic} skew scattering ($sk$),  \textit{extrinsic} side-jump ($sj$), and \textit{intrinsic} ($int$) parts~\cite{sinova2015,nagaosa2010,chadova2} as $\sigma_{xy}^s=\theta^{sk,s} \sigma_{xx}^s+\sigma_{xy}^{sj,s}+\sigma_{xy}^{int,s}$ giving thus the expression of the AHE in the conductivity tensor by summing the two spin-channel contributions $\sigma_{xy}=\theta^{sk} \sigma_{xx}+\sigma_{xy}^{sj}+\sigma_{xy}^{int}$ with $\theta^{sk}=\frac{\theta^{sk,\uparrow}\sigma_{xx}^\uparrow+\theta^{sk,\downarrow} \sigma_{xx}^\downarrow}{\sigma_{xx}^\uparrow+\sigma_{xx}^\downarrow}$, $\sigma_{xy}^{sj}=\sigma_{xy}^{sj,\uparrow}+\sigma_{xy}^{sj,\downarrow}$ and $\sigma_{xy}^{int}=\sigma_{xy}^{int,\uparrow}+\sigma_{xy}^{int,\downarrow}$. The rule to sum the different contributions to the off-diagonal transverse components for the conductivity tensors, \textit{i.e.} for different series processes in the electronic transport, originates from the same form of the off-diagonal components of the resistivity tensor:
\begin{eqnarray}
-\rho_{xy}^s=\theta^{sk,s} \rho_{xx}^s+\left(\sigma_{xy}^{sj,s}+\sigma_{xy}^{int,s}\right) \left(\rho_{xx}^{s~2}+\rho_{xy}^{s~2}\right)
\end{eqnarray}
where $\theta^{sk,\uparrow\downarrow}$, $\sigma_{xy}^{sj,\uparrow\downarrow}$ and $\sigma_{xy}^{int,\uparrow\downarrow}$ may be considered as independent of the resistivity of the host material which makes such formula very general and useful in practical cases~\cite{casanova2018}. If one considers only extrinsic skew-scattering SHE/AHE processes~\cite{fert1981}, one finally gets $\rho_{AHE}=-\left(\rho_{xy}^\uparrow\frac{\rho_{xx}^2}{\rho_{xx}^{\uparrow~2}}+\rho_{xy}^\downarrow\frac{\rho_{xx}^2}{\rho_{xx}^{\downarrow~2}}\right)$
like proposed in Ref.~\cite{fert3} giving \textit{in fine}~\cite{fert1981}:

\begin{equation}
-\rho_{AHE}=\left(\frac{\rho_{xy}^\uparrow+\rho_{xy}^\downarrow}{4}\right)\left(1+\mathcal{P}^2(z)\right)+\left(\frac{\rho_{xy}^\uparrow-\rho_{xy}^\downarrow}{2}\right)\mathcal{P}
\label{rhoahe}
\end{equation}
with the local spin-polarization $\mathcal{P}(z)$ within the multilayer along the coordinate $z$ writes $\mathcal{P}(z)=\frac{\sigma_{xx}^\uparrow(z)-\sigma_{xx}^\downarrow(z)}{\sigma_{xx}^\uparrow(z)+\sigma_{xx}^\downarrow(z)}$ is the spin-polarization $\mathcal{P}(z)$ that we have to calculate with our numerical procedure. The partition of the AHE resistivity according to the formula [~\ref{rhoahe}] was discussed by Fert \textit{et al.} as the partition between the \textit{spin-effect} (first term on the right-hand side) and the \textit{pure skew-scattering} effect (second term on the right-hand side). Such partition is also discussed in term of \textit{SAHE} (spin anomalous Hall effect) and \textit{SHE} for a magnetization control of the spin-currents in ferromagnets\cite{amin2019}. This second term (SHE) is also present for non magnetic materials giving rise to the standard SHE as far as  $\rho_{xy}^\downarrow-\rho_{xy}^\uparrow$ when the time-reversal symmetry may act (no magnetism).

\subsection{AHE in multilayers}

Our analyses is based on a semi-phenomenological approach of spin-dependent transport, involving spin-dependent diffusion and electron scattering at the multiple interfaces, possibly considering spin-flips caused by the local spin-orbit interactions. The principle of the method is then:

1)\qquad a representation of the spin-current profiles within the whole multilayers by correctly matching the out-of-equilibrium
electronic Fermi distributions. This is of \textcolor{black}{primary} importance to provide the amplitude of the anomalous Hall effect (AHE) in multilayers as far as a
ferromagnetic layer ensures the polarization of the current \textit{via} the non-local spin-dependent conductivity calculated by this method.

2)\qquad The adjacent SOC layer (Pt, Au:W) can affect the spin orbit assisted diffusion of electrons by \textit{spin-current proximity effect}. Therefore, the
AHE may be enhanced by increasing the spin orbit coupling in SOC layer. The more important is the SOC, the largest is the transverse spin current \textit{via}
the spin-to-charge conversion process and the Anomalous Hall angle. Eventhough, AHE may inverse its sign by changing the sign of SOC (Pt/CoNi, even Au:W$_{80}$).

\vspace{0.1in}

In the case of multilayers, non-local spin-dependent conductivities are necessary to consider leading to important spin-current proximity effects. Consequently theory beyond standard linear theory response of bulk materials like described here are necessary to correctly interprete the data. Our semi-phenomenological approach of spin-orbit polarized transport in magnetic multilayers, presented here, is based on a generalization of the theory of Camley-Barnas~\cite{camley1989,camley1990,ohno2020}. The counterpart of the relative simplicity of such semi-classical modelling is to find the appropriate boundary conditions to be adopted at interfaces like the ones proposed in Ref.~\cite{butler1995}. Those are given by a certain continuity/discontinuity relation of the out-of-equilibrium distributions of the functions of Fermi \textit{vs.} the transmission/reflection coefficients of the electronic waves and another coefficient of specularity $p$. This coefficient of specularity $p$ reflects the effects of 'isotropic' diffusions on the impurities at the interfaces and therefore plays a major role in particular on the properties of electronic reflection. Adapting the Camley-Barnas model to the model of Kubo-Greenwood's linear response in multilayers therefore requires finding the correct semi-phenomenological parameter set at the interfaces in order to fairly describe the diffusion properties.

\subsubsection{general overview.}

In the current-in-plane geometry (CIP) geometry, spin-polarized currents are translational invariant along the flow direction leading to a divergent-less current by construction with the results that no charge nor spin-accumulation occurs. The only first term to consider is the displacement of the spin-polarized currents on the Fermi surface. The counter-part for CIP currents is that one cannot define a single out-of-equilibrium length of the Fermi distribution near each
interfaces, being scaled for each direction by a $\cos(\theta)$ feature where $\theta$ represents the angle of incidence with respect to the
multilayered surface. In a semi-classical approach, CIP can be described through the well parametrized Boltzmann theory adapted to spin-polarized
transport according to the following general equation applied to the out-of-equilibrium Fermi distribution $f_{s}(\mathbf{r},\mathbf{v},\mathbf{E,}t)$ of spin $s$ under the application of an electric field $\mathbf{E}$ along the $z$ direction perpendicular to the layers according to:

\begin{equation}
f^s(z,\mathbf{v})=f_{0}(\mathbf{v})+g^{s}(z,\mathbf{v}).  \label{f1}
\end{equation}
where we remind that $e=|e|$ is absolute value of the electron charge, $\mathbf{v}$ is electron velocity, and where the distribution function for electron is decomposed into two parts; the equilibrium in zero electric field distribution function $f_{0}(\mathbf{v})$ and a small perturbation $g^{s}(z,\mathbf{v})$ induced by the
electric field and the interface scattering perturbation. The general solution $g^{s}(z,\mathbf{v})$ in each layer reads now:
\begin{equation}
g^{s}(z,\mathbf{v})=\left\{
\begin{array}{c}
\frac{eE\tau ^{s}}{m}\frac{\partial f_{0}(\mathbf{v})}{\partial v_{x}}\left(1+F_{+}^{s}(\mathbf{v})\exp \left( \frac{-z}{\tau ^{s}\left\vert
v_{z}\right\vert }\right) \right) \text{ for }v_{z}>0 \\
\frac{eE\tau ^{s}}{m}\frac{\partial f_{0}(\mathbf{v})}{\partial v_{x}}\left(1+F_{-}^{s}(\mathbf{v})\exp \left( \frac{z}{\tau ^{s}\left\vert
v_{z}\right\vert }\right) \right) \text{ for }v_{z}<0
\end{array}
\right.  \label{g1}
\end{equation}%
where $F_{\pm }^{s}(\mathbf{v})$ are arbitrary functions of the electron velocity and found by the matching conditions. The current density for the spin $s$ is given by:
\begin{equation}
J^{s}(z)=-2|e|\left( \frac{m}{\hbar }\right) ^{3}\int_{V}v_{x}g_{x}^{s}(z,\mathbf{v})d\mathbf{v,}  \label{js}
\end{equation}%
where $V$ is the unit volume, $d\mathbf{v=}dv_{x}dv_{y}dv_{z}$ represent the 3D velocity space, and $g^{s}(z,\mathbf{v})$ may be decomposed into a bulk term $g^{s}_0$, plus an out-of-equilibrium interface term $\delta g^{s}$ and originating from interface scattering with the result that $g^{s}=g^{s}_0+\delta g^{s}$.

\subsubsection{treatment in multilayers.}

The whole matching conditions for the out-of-equilibrium Fermi distribution $g^s$ to fulfill at interfaces are considered \textit{via} the S-scattering matrix formalism. $g_{0}^{s}$ and $\delta g^{s}$  describe respectively the the bulk conduction and the perturbation originating from the interface terms due to interface scattering (reflection/transmission). In a spinor form, we have $g_{0}=\left[
\begin{array}{cc}
g_{0}^{\uparrow} ,& g_{0}^{\downarrow}
\end{array}%
\right]^T,$ and $\delta g_{0}=\left[
\begin{array}{cc}
\delta g^{\uparrow} ,& \delta g^{\downarrow}%
\end{array}%
\right]^T$ to give $g=\left[
\begin{array}{cc}
g^{\uparrow} ,& g^{\downarrow}
\end{array}%
\right]^T =\left[
\begin{array}{cc}
g_{0}^{\uparrow }+\delta g^{\uparrow} ,&
g_{0}^{\downarrow }+\delta g^{\downarrow}
\end{array}%
\right]^T$, $g^{+\text{ }}$ for electrons possessing a positive velocity $v_{z}$ along $z$ and $g^{-\text{}}$ for carriers with opposite velocity $-v_{z\text{{}}}$.

\vspace{0.1in}

If one considers the scattering at a single interface, and before generalizing to multilayers, it may be written as:
\begin{equation}
\left(
\begin{array}{c}
g_{L}^{-} \\
g_{R}^{+}%
\end{array}%
\right)=\left(
\begin{array}{cc}
R & T^{\prime} \\
T & R^{\prime}%
\end{array}%
\right) \left(
\begin{array}{c}
g_{L}^{+} \\
g_{R}^{-}%
\end{array}%
\right),  \label{0.1}
\end{equation}%
where $g_{L}^{\pm }$ and $g_{R}^{\pm }$ represent the respective distribution functions to the left and to the right of the interface, $T,R$ are the respective transmission and reflection spin-dependent transmission coefficients from the left to the right; whereas $T^{\prime },R^{\prime}$ are the ones from the right to the left. Transmission and reflection coefficients are built in a 2$\times 2$ matrix form, with \textit{e.~g.} $R=\left(
\begin{array}{cc}
R^{\uparrow \uparrow } & R^{\uparrow \downarrow } \\
R^{\downarrow \uparrow } & R^{\downarrow \downarrow }%
\end{array}%
\right) $ and $T=\left(
\begin{array}{cc}
T^{\uparrow \uparrow } & T^{\uparrow \downarrow } \\
T^{\downarrow \uparrow } & T^{\downarrow \downarrow }%
\end{array}%
\right)$ where the respective $\uparrow \uparrow$ and $\downarrow \downarrow$ stands for the spin-conserving terms wheres $\uparrow \downarrow$ and $\downarrow \uparrow$ for the spin-flip terms. It turns that:

\begin{equation}
T=\left[
\begin{array}{cc}
T^{\uparrow }\left( 1-\frac{sf}{2}\right) ^{2}+T^{\downarrow }\left( \frac{sf%
}{2}\right) ^{2} & \left( T^{\uparrow }+T^{\downarrow }\right) \frac{sf}{2}%
\left( 1-\frac{sf}{2}\right) \\
\left( T^{\uparrow }+T^{\downarrow }\right) \frac{sf}{2}\left( 1-\frac{sf}{2}%
\right) & T^{\downarrow }\left( 1-\frac{sf}{2}\right) ^{2}+T^{\uparrow
}\left( \frac{sf}{2}\right) ^{2}%
\end{array}%
\right]  \label{0.1bis}
\end{equation}%
where $T^{\uparrow}=\frac{T^*}{1-\gamma}$, $T^{\downarrow }=\frac{T^*}{1+\gamma }$ and $T^*$ is a certain average of the transmission, $\gamma$ is the interfacial spin-transmission asymmetry parameter, $(sf)=1-\exp{-\delta}$ is the interfacial spin-flip probability and $\delta$ the spin memory loss parameter~\cite{bass}. The same relationship exist for the reflection matrix $R$. The degree of specularity at interfaces, $sp^{R}$ and $sp^{T}$, in the respective reflection/transmission processes may be taken into account \textit{via} $R=R_{sp=1}\times sp^{R}$ and $T=T_{sp=1}\times sp^{T}$ where $sp=1$ means that $R$ and $T$ correspond to full specular processes.

\vspace{0.1in}

\emph{case of a single interface}

\vspace{0.1in}

The case of a single interface in then treated by the following relationship linking left and right out-of-equilibrium components:
\begin{eqnarray}
\left(
\begin{array}{c}
1_{2\times 1} \\
\delta g_{L}^{-} \\
\delta g_{R}^{+}%
\end{array}%
\right) =\left(
\begin{array}{ccc}
1_{2\times 2} & 0_{2\times 2} & 0_{2\times 2} \\
\Sigma^{(+)} & R & T^{\prime } \\
\Sigma^{(-)} & T & R^{\prime }%
\end{array}%
\right) \left(
\begin{array}{c}
1_{2\times 1} \\
\delta g_{L}^{+} \\
\delta g_{R}^{-}%
\end{array}%
\right)
\end{eqnarray}
where $1_{2\times 1}=\binom{1}{1}$, $1_{2\times 2}=\left(
\begin{array}{cc}
1 & 0 \\
0 & 0%
\end{array}%
\right) ,$ $0_{2\times 2}=\left(
\begin{array}{cc}
0 & 0 \\
0 & 0%
\end{array}%
\right)$.
with the source terms $\Sigma^{(+)}$ and $\Sigma^{(-)}$ written in a 2x2 diagonal matrix form:
\begin{eqnarray}
\Sigma^{(+)}=\left[\left(R-1\right) g_{0L}+T^{\prime}g_{0R}\right]\\
\Sigma^{(-)}=\left[Tg_{0L}+\left(R^{\prime}-1\right) g_{0R}\right]
\end{eqnarray}
where we have considered the respective \textit{left-travelling} ($-$) and \textit{right travelling} ($+$) electrons and their bulk out-of-equilibrium Fermi distributions, $g_{0L}^+=g_{0L}^-=g_{0L}$ and $g_{0R}^+=g_{0R}^-=g_{0R}$, at the respective left ($L$) and right ($R$) side of the interface. The following $S_{6\times 6}$ matrix is revealed to be the relevant scattering matrix to find the interfacial out-of-equilibrium distribution function:

\begin{equation}
S=\left(
\begin{array}{ccc}
1_{2\times 2} & 0_{2\times 2} & 0_{2\times 2} \\
\Sigma^{(+)} & R & T^{\prime } \\
\Sigma^{(-)} & T & R^{\prime }%
\end{array}%
\right) .  \label{S}
\end{equation}%
before generating it by recursion to the case of multiple interfaces.

\vspace{0.1in}

\emph{case of multiple interfaces: the scattering path method.}

\vspace{0.1in}

One has to define here each propagation matrix $P^{(i)}$ inside a given layer $(i)$ like $P^{(i)}=\text{Diag}_{2\times2}\left[\exp\{-\frac{d(i)}{\lambda^{(i)\uparrow}\cos\theta}\};\exp\{-\frac{d(i)}{\lambda^{(i)\downarrow}\cos\theta}\}\right]$ where $d^{(i)}$ is layer thickness of the corresponding layer and $\lambda^{(i)\uparrow \left(\downarrow \right)}$ is the mean free path of the respective $\uparrow, \downarrow $ spin. One uses then the scattering-path method applied to the out-of-equilibrium Fermi distribution functions $g$ to solve the entire problem. This method consists in searching for the solution of the interfacial out-of-equilibrium contribution function $\delta g_{j}^{\pm}$ (here $j$ labels the interface) according to:

\begin{equation}
\hat{g}_{out}^{(i)}=\Sigma_{(j)}\hat{S}^{(ij)}~\hat{g}_{int}^{(j)}  \label{super matrix relation}
\end{equation}%
written in a super-matrix form. The 'supermatrix' denomination means that, on the above equation, we apply a double summation made on the two-dimensional spinor space (spin-conserving and spin-flip terms) as well on the (j)-space using twice the Einstein's notation. $\hat{S}$ is a $6N\times 6N$ matrix size with $N$ the number of interfaces. It has the meaning of finding the correlation between transmitted and reflected out-of-equilibrium function distribution at the surface (i) from \textit{perturbations} generated from the interface (j) and where
\begin{equation}
\hat{g}_{int}=\left[
\begin{array}{c c c c}
g_{int}^{(1)} & g_{int}^{(2)} & .. & g_{int}^{(N)}
\end{array}\right]^T
\label{int}
\end{equation}
with $g_{int}^{(N)}$ a column vector with $g_{int}^{(1)}=\left(
\begin{array}{ccc}
1_{_{2\times 1}} & 0_{2\times 1} & \delta g_{(1)}^{-}
\end{array}\right)^T,\qquad
g_{int}^{(N)}=\left(
\begin{array}{ccc}
1_{_{2\times 1}} & \delta g_{N+1}^{+} & 0_{2\times 1}
\end{array}\right)^T$ and
$g_{int}^{(n)\neq \left(1,N\right)}=\left(
\begin{array}{ccc}
1_{_{2\times 1}} & 0 & 0
\end{array}\right)^T$ is the known $6\times 1$ matrix incoming or source components whereas $\hat{g}_{out}$ is given by:

\begin{equation}
\hat{g}_{out}^{(i)}=\left[
\begin{array}{ccccccc}
1_{2\times 1} & \delta g_{(i)}^{-} & \delta g_{(i+1)}^{+}
\end{array}
\right]^T  \label{out}
\end{equation}

One may then consider the general relationship linking the generalized scattering matrix $S_{nm}$ (with \textit{large} S) to the respective single-interface scattering matrices $s_{n}$ at the interface $n$ according to:
\begin{eqnarray}
S_{(ij)}=s_i\delta_{ij}-s_i~P_{(il)}~S_{(lj)}
\end{eqnarray}
or equivalently
\begin{eqnarray}
\left[S\right]^{-1}_{(ij)}=\left(\left[s_i\right]^{-1}\delta_{ij}-P_{(ij)}\right)
\end{eqnarray}
where $\delta_{ij}$ is the Kronecker symbol and where $i,j,l$ denote the interface and $P_{(il)}$ the propagator from the interface $i$ to the interface $j$. $s_{i}$ is the single-interface $6\times 6$ scattering matrix at the interface $(i)$. In particular, one has the self-consistent equation for each $(i)$:
\[
\left[
\begin{array}{ccc}
1_{2\times 1} & \delta g_{i}^{-} & \delta g_{i+1}^{+}%
\end{array}%
\right]^T =s_{i}\left[
\begin{array}{ccc}
1_{2\times 1} & \delta g_{i}^{+} & \delta g_{i+1}^{-}%
\end{array}%
\right]^T ,\]

with
\begin{eqnarray}
P_{n,n+1}=\left(
\begin{array}{ccc}
0_{2\times 2} & 0_{2\times 2} & 0_{2\times 2} \\
0_{2\times 2} & 0_{2\times 2} & 0_{2\times 2} \\
0_{2\times 2} & 1_{2\times 2} & 0_{2\times 2}\\
\end{array}
\right);
P_{n+1,n}=\left(
\begin{array}{ccc}
0_{2\times 2} & 0_{2\times 2} & 0_{2\times 2} \\
0_{2\times 2} & 0_{2\times 2} & 1_{2\times 2} \\
0_{2\times 2} & 0_{2\times 2} & 0_{2\times 2}%
\end{array}%
\right) \nonumber \\
\end{eqnarray} with $n\geq 1.$

\begin{figure}[tbp]
\includegraphics[width=8cm]{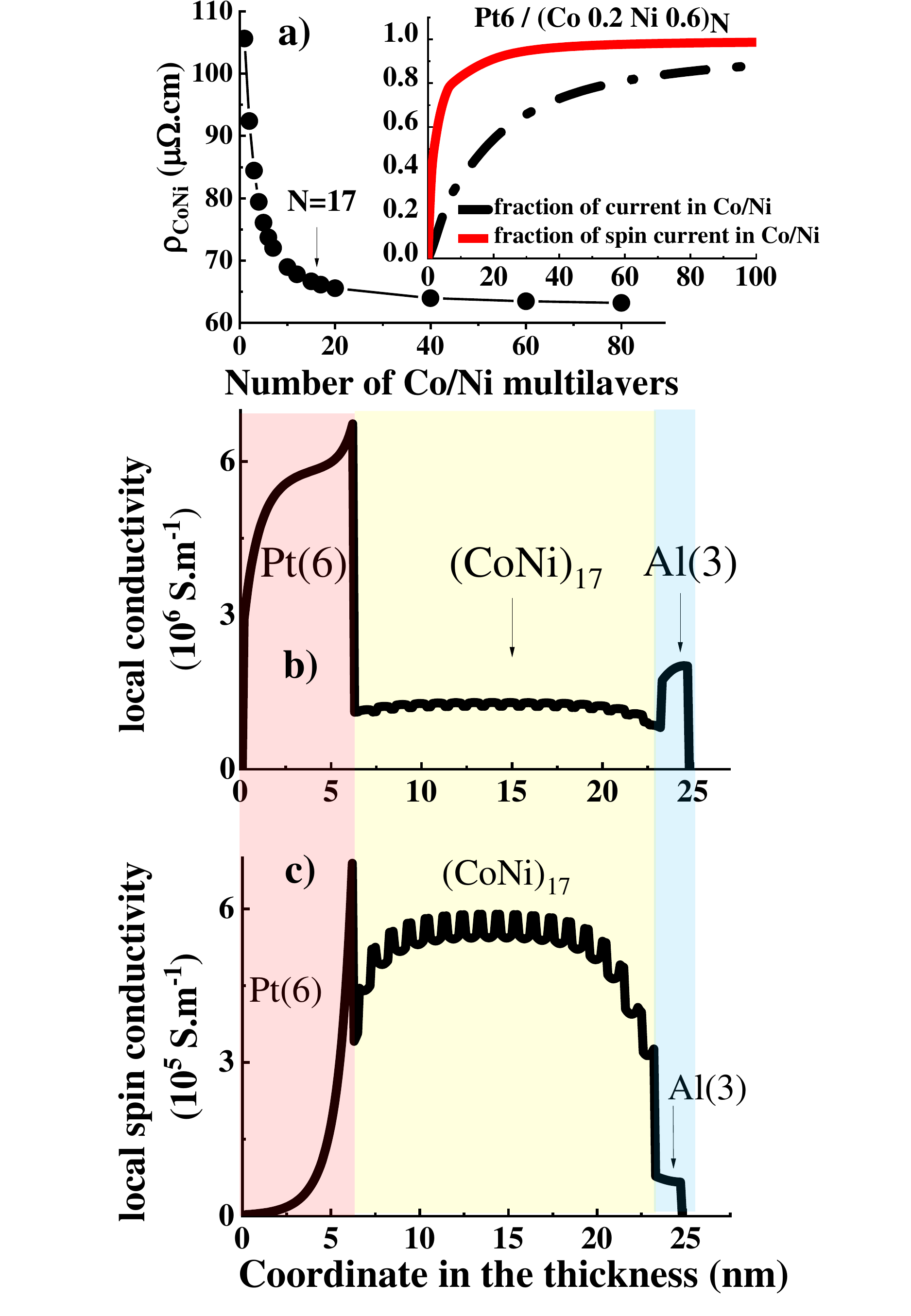}
\caption{a) Average resistivity of (Co0.2/Ni0.6)$_N$ \textit{vs.} number of (Co/Ni) repetitions induced by electronic interface scatterings. a inset) fraction of current and spin-current in (Co/Ni) multilayers in Pt(6)/(Co0.2/Ni0.6)$_N$ systems showing a spin-polarisation of Pt as the total thickness of (Co/Ni) is small ($N$ small). b) Local conductivity in the Pt(6)/[Co/Ni]$_{17}$ structure taking into account electronic interface scatterings. c) Local spin-conductivity in the Pt(6)/(Co/Ni)$_{17}$ structure taking into account electronic interface scatterings. The physical parameters are given in Table~\ref{TableI}.}
\label{fig:spincurrent}
\end{figure}

\vspace{0.1in}

\textcolor{black}{We have checked that our code was first able to simulate in a very good agreement the spin-current profiles and CIP-GMR from Co/Cu/NiFe structures like provided by ab-initio techniques~\cite{butler1995} and using known parameters for Co, Cu and NiFe extracted from the literature~\cite{supplemental}.} Examples of calculations of the properties of Pt/(Co,Ni) multilayered system \textcolor{black}{for} AHE is displayed on Fig.~\ref{fig:spincurrent} using the ensemble of physical parameters given in Table~\ref{TableI}. Fig.~\ref{fig:spincurrent}a displays the typical increase of the resistivity of the (Co0.2/Ni0.6)$_N$ system ($N$ is the number of repetitions) from the bulk value ($60 \mu\Omega.$cm) at RT up to $110 \mu\Omega.$cm for $N\simeq 1-2$ due to electronic scatterings and non-specularity at interfaces ($sp^R_{Co/Pt}=0,sp^R_{Ni/Al}=0$ and $sp^R_{Co/Ni}$=0.4) and in particular at the outer boundaries of the structure. In parallel, the inset of this figure displays the dependence in $N$ of the fraction of charge current and spin-currents in (Co/Ni). The complementary to the unit 1.0 gives the same information in Pt. The important information lies in the proportion of spin-current in Pt for $N$ small. A typical value of about 10-50\% of spin-current shunt in Pt for $N<17$ may lead to AHE sign inversion owing to a large value of SHA of Pt.

Fig.~\ref{fig:spincurrent}b-c display both the local longitudinal conductivity $\sigma_{xx}(z)$ (charge current density for unit Electric field) and local longitudinal spin-conductivity $\sigma_{xx}^z$ with magnetization along $z$ for the corresponding  Pt6/(Co0.2/Ni0.6)$_N$/Al3 structure (Fig.~\ref{fig:spincurrent}c). Those figures display clearly the shunt of the current in the Pt6 layer (Fig.~\ref{fig:spincurrent}b) owing to the higher conductivity of Pt compared to the thin-film (Co/Ni) layer together with the proximity \textit{leakage} spin-current in Pt and originating from (Co/Ni); the transmission coefficient at Co/Pt interface for the majority spin is set to unity with a full transmission specularity. The typical decrease of the spin-current in Pt within 1~nm scales with the electronic mean-free path. Such leakage spin-current in Pt is responsible for the unconventional inverted anomalous Hall effect due to opposite sign of the spin-orbit interaction between Pt and (Co/Ni) described here.

\section{Data analysis and interpretation}

We now focus on the experimental results and analyze quantitatively the data using our theoretical basis. How to explain such observations? The spin-current is generated in (Co/Ni) multilayers with a given spin-polarization from the ferromagnetic bulk properties. However, as the number of sequences $N$ for (Co/Ni) ($N=3-5$) remains small, the current partly spin-polarized, occurs to be dominant in Pt compared to the (Co/Ni) region of reduced thickness. This occurs up to a given threshold limit of $N$ above which the conduction becomes dominant in (Co/Ni) like in the case of Pt/(Co/Ni)$_{20}$ and Au:W$_{130}$/(Co/Ni)$_{40}$ (Fig.~\ref{fig:1}). The semi-phenomenological theory of current-in-plane (CIP) spin-currents~\cite{camley1989} indeed shows that (Co/Ni) is able to provide the necessary polarized current within the whole stacks including Pt (Fig.~\ref{fig:spincurrent}). The existence of such spin-polarized proximity current is converted into a transverse current \textit{via} the local SOI and ISHE. We thus demonstrate that Pt possesses a positive spin Hall angle~\cite{jaffres2014,laczkowski2017} while (Co/Ni) with thicker Ni possesses a negative SHE sign. This is corroborated by following modeling and simulations presented in that second part.

\subsection{Physical models for AHE.}

In order to retain the main physical principles driving the SHE and AHE in MLs with Pt related interfaces, our idea is to distinguish the four possible different mechanisms of AHE in (Co/Ni): \textit{(i)} an \textit{intrinsic} AHE-SHE phenomenon in (Co/Ni) viewed as an \textit{effective} material and characterized by an average spin-Hall conductivity (SHC=$\sigma_{xy}^{int,s}$), a pure \textit{extrinsic} SHE mechanism acting either on \textit{(ii)} the majority or \textit{(iii)} the minority spin channels with an overall \textit{extrinsic} SHA given by $\theta^{eff}=\frac{\theta^\uparrow\sigma^\uparrow+\theta^\downarrow\sigma^\downarrow}{\sigma^\uparrow+\sigma^\downarrow}$. A larger majority spin-current is expected ($\sigma^\uparrow> \sigma^\downarrow$) whereas a larger SHA is expected in the spin minority band ($|\theta^\uparrow|<|\theta^\downarrow|$) by enhanced \textit{sp-d} band mixing and necessary phase shift for skew-scattering phenomena~\cite{levy1988,maekawa2009}. This brings uncertainties between scenario \textit{(ii)} and \textit{(iii)} for extrinsic mechanism like suggested in Ref.~\cite{fert1981,zimmermann}. In that sense, our approach is slightly different from considering an identical SHA for both spin channel~\cite{casanova2018}. The last scenario \textit{(iv)} is the one of magnetic induced moment in Pt (MPE) generating spin-currents and AHE in Pt close to the Co interface at the scale of a few (typically 2) atomic planes~\cite{mokrousov2015,bailey2016,crowell2018}.

Apart from spin-dependent electronic diffusions in bulk, one may emphasize the relevant boundary conditions to match for the out-of-equilibrium Fermi distribution in the framework of Fuchs-Sondheimer model~\cite{fuchssondheimer}. This is generally performed by including possible specular~\cite{stewart2003,chen17} or diffusive electron reflection (R)/transmission(T) at interfaces in the CIP spin-dependent Boltzmann equations involving layer- and spin-dependent electronic mean free path $\lambda_i^s$. One also has to consider the corresponding SOI spin-mixing terms in a $2\times 2$ Pauli matrix form and related spin-flip probability~\cite{heers2012}. This is particularly true at the Co/Pt interface where the spin-loss is known to be large. It is parameterized, here, by a spin-flip coefficient $p_{sf}$ related to the spin-memory loss (SML) $\delta $ parameter~\cite{jaffres2014} according to $p_{sf}=1-\exp(-\delta)$. SML at $3d-5d$ interfaces plays unavoidable role in spin-pumping in FMR experiments~\cite{jaffres2014,nist2018}. Moreover, one introduces the overall longitudinal resistivity $\rho _{xx}^{\ast}$ (or conductivity $\sigma_{xx}^{\ast}$) and transverse resistivities $\rho _{xy}^{\ast}$ (or transverse conductivity $\sigma_{xy}^{\ast}$) of the MLs as:

\begin{eqnarray}
R_{xx}=\rho _{xx}^{\ast}\frac{L}{Wt}\simeq\frac{L}{W}\frac{1}{t\sigma_{xx}^{\ast }}=\frac{L}{W}\frac{1}{\sum_{i,s}\sigma _{xx,i}^st_{i}},\\
R_{xy}\simeq\frac{\rho _{xy}^{\ast }}{t}=\frac{\sigma _{xy}^{\ast }}{t\left(\sigma _{xx}^{\ast}\right) ^{2}}=\frac{\sum_{i,s}\left(\sigma_{xy,i}^{s}t_i
\right)}{\left(\sum_{i,s}\sigma_{xx,i}^s t_i\right)^2}
\label{transverse resistance}
\end{eqnarray}
where $L,$ $W$ represents the length and width of the Hall cross bars, $t$ is the overall thickness of MLs and $\sigma _{xx,i}^s$ the local
longitudinal spin-conductivity of the $i^{th}$ layer of thickness $t_{i}$ and $\sigma_{xy,i}^s$ the local off-diagonal spin-conductivity in the layer $i$.

\begin{table*}
\begin{center}
\scalebox{0.8}{
\begin{tabular}{|l||l||l|l|}
\hline
Parameters & Symbols & Values for $\uparrow $ spin & Value for $\downarrow$ spin \\ \hline \hline
Conductivity of Co & $\sigma _{Co}\left( S.m^{-1}\right) $ & \multicolumn{1}{|c|}{$\frac{\sigma _{Co}}{1-\beta _{Co}}=7.4\times 10^{6}$} & \multicolumn{1}{|c|}{$\frac{\sigma _{Co}}{1+\beta _{Co}}=2.7\times 10^{6}$} \\ \hline
Conductivity of Ni & $\sigma _{Ni}\left( S.m^{-1}\right) $ & \multicolumn{1}{|c|}{$\frac{\sigma _{Ni}}{1-\beta _{Ni}}$=$1.7\times 10^{7}$} & \multicolumn{1}{|c|}{$\frac{\sigma _{Ni}}{1+\beta _{Ni}}=2.3\times 10^{6}$} \\ \hline
Conductivity of Pt & $\sigma _{Pt}\left( S.m^{-1}\right) $ & \multicolumn{1}{|c|}{$4.5\times 10^{6}$} & \multicolumn{1}{|c|}{$4.5\times 10^{6}$} \\ \hline
Conductivity of Au:W$_{130}$ & $\sigma _{Au:W_{130}}\left(S.m^{-1}\right) $ & \multicolumn{1}{|c|}{$3.85\times 10^{5}$} & \multicolumn{1}{|c|}{$3.85\times 10^{5}$} \\ \hline
Conductivity of Au:W$_{80}$ & $\sigma _{Au:W_{80}}\left( S.m^{-1}\right) $ & \multicolumn{1}{|c|}{$6.2\times 10^{5}$} & \multicolumn{1}{|c|}{$6.2\times 10^{5}$} \\ \hline \hline\\
Intrinsic Hall conductivity of Co/Ni (case \textit{(i)}) & $\sigma_{xy}^{int}\left( S.cm^{-1}\right) $ & \multicolumn{2}{|c|}{$=\sigma_{xy}^{int,\uparrow }+\sigma _{xy}^{int,\downarrow }=-85$} \\ \hline \hline
Mean free path of Co & $\lambda _{Co}(nm)$ & \multicolumn{1}{|c|}{$7.4$} & \multicolumn{1}{|c|}{$2.7$} \\ \hline
Mean free path of Ni & $\lambda _{Ni}(nm)$ & \multicolumn{1}{|c|}{$16$} & \multicolumn{1}{|c|}{$2.3$} \\ \hline
Mean free path of Pt & $\lambda _{Pt}(nm)$ & \multicolumn{1}{|c|}{$1.6$} & \multicolumn{1}{|c|}{$1.6$} \\ \hline
Mean free path of Au:W & $\lambda _{Pt}(nm)$ & \multicolumn{1}{|c|}{$0.4$} & \multicolumn{1}{|c|}{$0.4$} \\ \hline \hline
Bulk asymmetry coefficient of Co & $\beta _{Co}$ & \multicolumn{2}{|c|}{$\ \ 0.46$} \\ \hline
Bulk asymmetry coefficient of Ni & $\beta _{Ni}$ & \multicolumn{2}{|c|}{$\ \ 0.76$} \\ \hline
Spin Hall angle of Co/Ni (case \textit{(ii)}) & $\theta _{Co/Ni}^{\uparrow }$ & \multicolumn{1}{|c|}{$-0.9\%$} & \multicolumn{1}{|c|}{$0$} \\ \hline
Spin Hall angle of Co/Ni (case \textit{(iii)}) & $\theta_{Co/Ni}^{\downarrow }$ & \multicolumn{1}{|c|}{$0$} & \multicolumn{1}{|c|}{$2.2\%$} \\ \hline
Spin Hall angle of Co/Ni (case \textit{(iv)}) & $\theta _{Co/Ni}^{{}}$ & \multicolumn{1}{|c|}{$-1.5\%$} & \multicolumn{1}{|c|}{$1.5\%$} \\ \hline
Spin Hall angle of Pt & $\theta _{Pt}^{{}}$ & \multicolumn{1}{|c|}{$20\%$} & \multicolumn{1}{|c|}{$-20\%$} \\ \hline
Spin Hall angle of Au:W$_{130}$ & $\theta _{Au:W_{130}}$ & \multicolumn{1}{|c|}{$-0.3\%$} & \multicolumn{1}{|c|}{$0.3\%$} \\ \hline
Spin Hall angle of Au:W$_{80}$ & $\theta _{Au:W_{130}}$ & \multicolumn{1}{|c|}{$10\%$} & \multicolumn{1}{|c|}{$-10\%$} \\ \hline \hline
Pt/Co Average interface transmission & $t_{Pt/Co}$ & \multicolumn{1}{|c|}{$\frac{t_{Pt/Co}}{1-\gamma _{Pt/Co}}=0.94$} & \multicolumn{1}{|c|}{$\frac{t_{Pt/Co}}{1+\gamma _{Pt/Co}}=0.31$} \\ \hline
Co/Ni Average interface transmission & $t_{Co/Ni}$ & \multicolumn{1}{|c|}{$\frac{t_{Co/Ni}}{1-\gamma _{Co/Ni}}=0.14$} & \multicolumn{1}{|c|}{$\frac{t_{Co/Ni}}{1+\gamma _{Co/Ni}}=0.08$} \\ \hline
Au:W/Co Average interface transmission & $t_{Au:W/Co}$ & \multicolumn{1}{|c|}{$\frac{t_{Au:W/Co}}{1-\gamma _{Au:W/Co}}=0.43$} & \multicolumn{1}{|c|}{$\frac{t_{Au:W/Co}}{1+\gamma _{Au:W/Co}}=0.23$} \\ \hline \hline
Pt/Co Interface asymmetry coefficient  & $\gamma _{Pt/Co}$ & \multicolumn{2}{|c|}{$0.5$} \\ \hline
Co/Ni Interface asymmetry coefficient  & $\gamma _{Co/Ni}$ & \multicolumn{2}{|c|}{$0.9$} \\ \hline
Au:W/Co  Interface asymmetry coefficient & $\gamma _{Au:W/Co}$ & \multicolumn{2}{|c|}{$0.3$} \\ \hline \hline
Pt/Co specularity in reflection & $sp_{Pt/Co}$ & \multicolumn{2}{|c|}{$0$}\\ \hline
Au:W/Co specularity in reflection & $sp_{Au:W/Co}$ & \multicolumn{2}{|c|}{$0$} \\ \hline
Co/Ni specularity in reflection & $sp_{Co/Ni}$ & \multicolumn{2}{|c|}{$0.4$} \\
\hline
Ni/Al specularity in reflection & $sp_{Co/Ni}$ & \multicolumn{2}{|c|}{$0$} \\
\hline \hline
Pt/Co spin-loss memory & $\delta _{Pt/Co}$ & \multicolumn{2}{|c|}{$0.9$} \\ \hline
Co/Ni spin-loss memory & $\delta _{Co/Ni}$ & \multicolumn{2}{|c|}{$0.3$} \\ \hline
Au:W/Co spin-loss memory & $\delta _{Au:W/Co}$ & \multicolumn{2}{|c|}{$0$}\\ \hline
\end{tabular}}
\caption{Table of physical parameters of Pt, Co and Ni extracted for Type II samples from our fit procedure in the case of pure \textit{intrinsic} and \textit{extrinsic} models for AHE. The error bar corresponds to the value of the latest significant figure.}
\label{TableI}
\end{center}
\end{table*}

\begin{figure}[tbp]
\includegraphics[width=9cm]{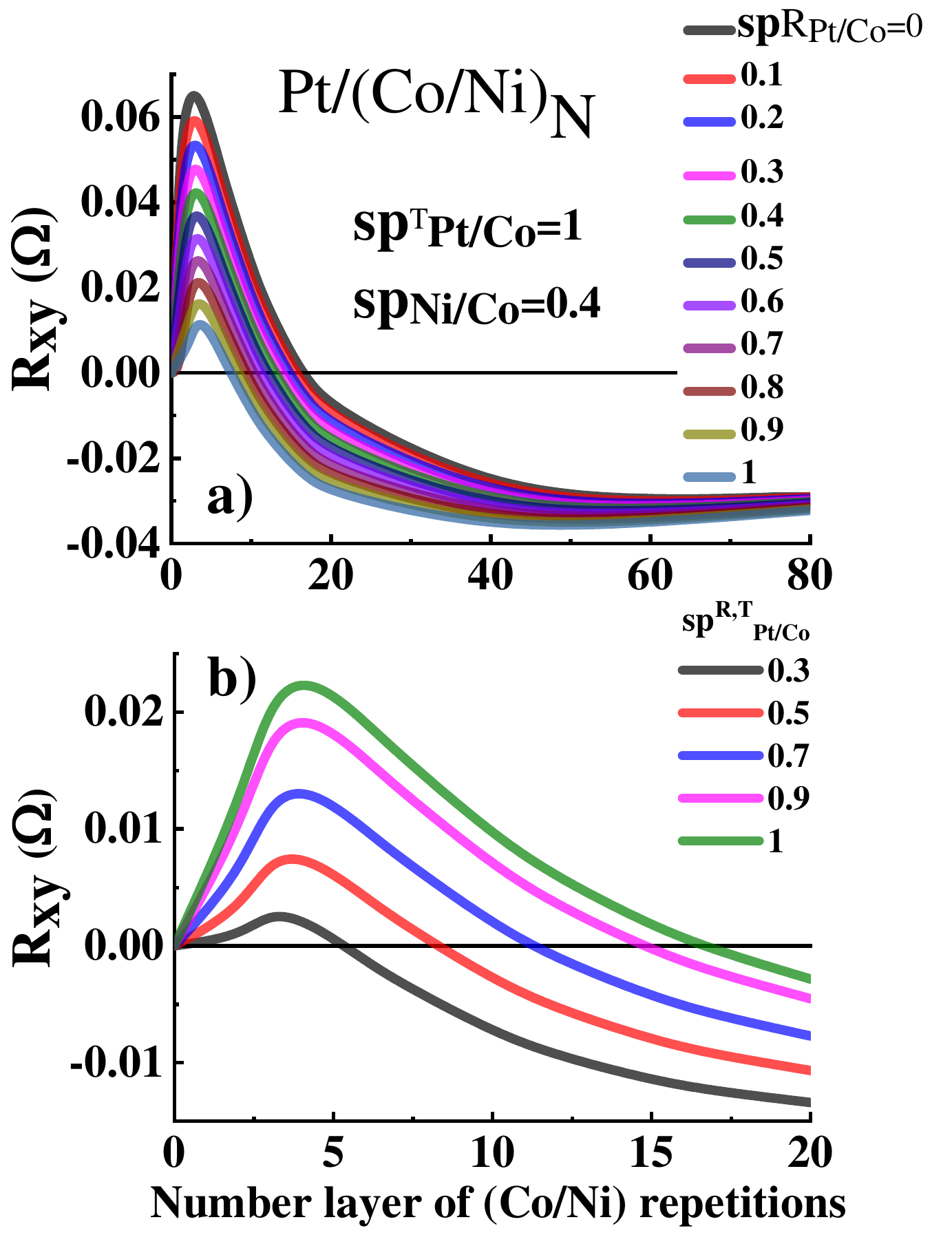}
\caption{a) Anomalous Hall resistivity $R_{xy}$ of (Co/Ni) in Pt(6)/(Co/Ni)$_N$ \textit{vs.} number of repetitions $N$ calculated for different specularity of electron reflection at (Co/Pt) interfaces. The specularity of transmission at  Co/Pt interface is set to 1, $sp_{Co/Pt}^T=1$ whereas at the $(Co/Ni)$ interface the values are fixed $sp_{Co/Ni}^T=sp_{Co/Ni}^R=0.4$. b) Anomalous Hall resistivity of Pt(6)/(Co/Ni)$_N$ \textit{vs.} number of repetitions $N$ of (Co/Ni) calculated for different specularities of reflection and transmission at (Co/Pt) interfaces $sp_{Co/Pt}^T=sp_{Co/Pt}^R$}.
\label{fig:specularity}
\end{figure}

\subsection{Effects of specularity on AHE.}

We address now in details the effect of specularity on the unconventional AHE of Pt6/(Co/Ni)$_N$ structures with the help of our theory and calculations. Fig.~\ref{fig:specularity} displays main of our results. We first discuss on Fig.~\ref{fig:specularity}a the effect of the specularity in reflexion $sp_{Co/Pt}^R$ of the Co/Pt interface, the transmission being fully specular as emphasized in Ref.~\cite{stewart2003}; the remaining parameters chosen for the simulations are given in Table~\ref{TableI}. One notes a clear decrease of the AHE resistivity $R_{xy}$ as $sp_{Co/Pt}^R$ increases from 0 to 1. This feature has to be associated to an increase of the spin-current in (Co/Ni) only leading in parallel to a better compensation of the spin-charge conversion between (Co/Ni) and Pt at small (Co/Ni) thickness thus leading to a crossover from positive to negative AHE occurring for a smaller number of repetition $N$.

\begin{figure}[tbp]
\includegraphics[width=8cm]{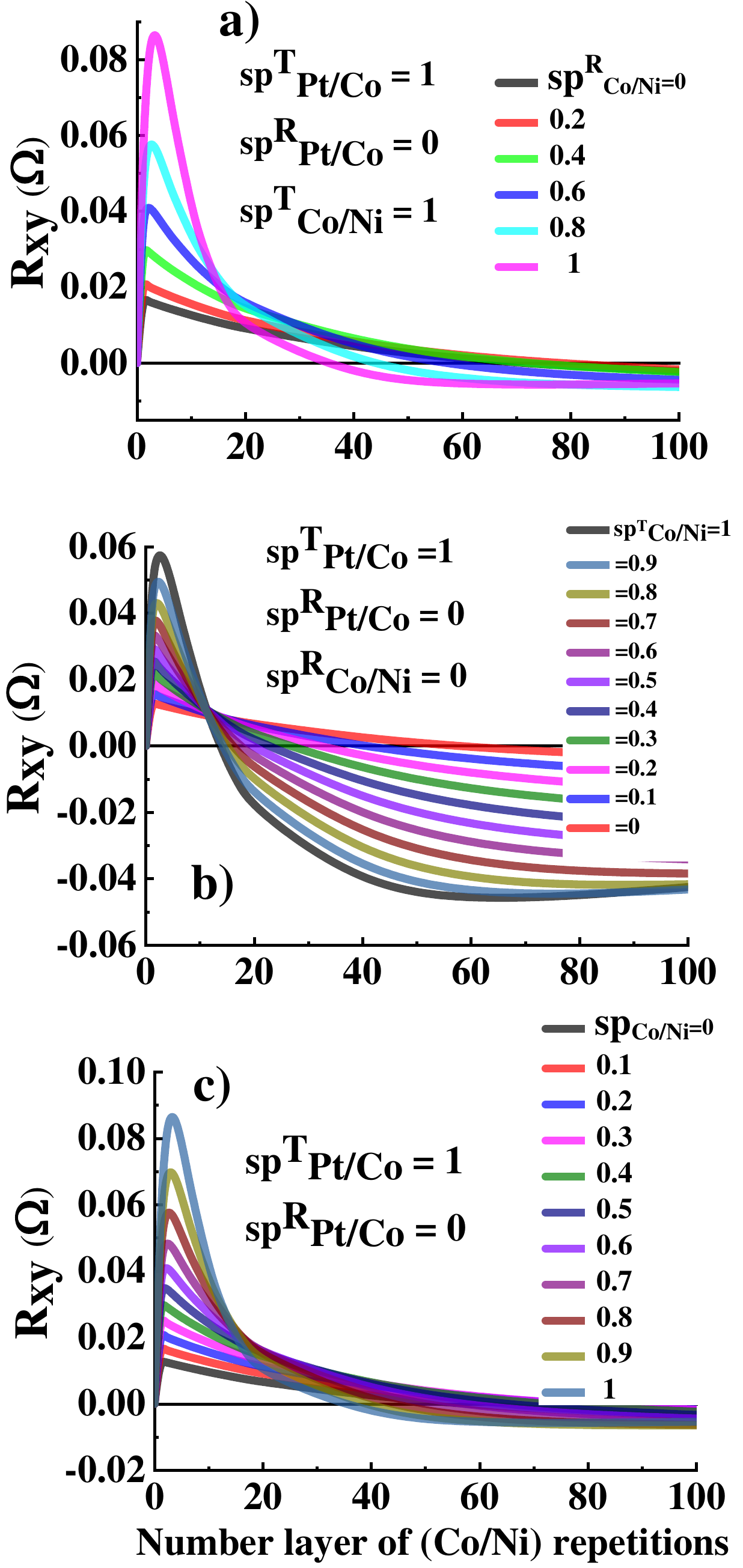}
\caption{a) Anomalous Hall resistivity $R_{xy}$ \textit{vs.} number of repetitions $N$ of (Co/Ni) in Pt(6)/(Co/Ni)$_N$ calculated for different specularity of electron reflection at (Co/Ni) interfaces. The specularity for transmission at Co/Pt (\textit{resp.} reflection) and Co/Ni interfaces is set to 1 (\textit{resp.} 0). b) Anomalous Hall resistivity \textit{vs.} number of repetitions $N$ of (Co/Ni) in Pt(6)/(Co/Ni)$_N$ calculated for different specularity of transmission at (Co/Ni) interfaces. The specularity at reflection at Co/Ni interfaces is set to 0 whereas the specularity at transmission for the Co/Pt interface is set to 1. c) Anomalous Hall resistivity \textit{vs.} number of repetitions $N$ of (Co/Ni) in Pt(6)/(Co/Ni)$_N$ for different joint variations of specularity at reflection and transmission for Co/Ni $sp_{Co/Pt}^R=sp_{Co/Pt}^T$.}
\label{fig:specularity_CoNi}
\end{figure}

On Fig.~\ref{fig:specularity}b, we have changed both the specularity in reflection $sp_{Co/Pt}^R$ and in transmission $sp_{Co/Pt}^T$ at the Pt/Co interface in the same manner ($sp_{Co/Pt}^R=sp_{Co/Pt}^T$) from 0.3 to 1. Compared to the previous case where $sp_{Co/Pt}^T=1$ and $sp_{Co/Pt}^R=0$, one can observe that \textit{i)} the $R_{xy}$ signal drops from 0.06~$\Omega$ to 0.02~$\Omega$ for its maximal value clearly indicating a reduction of the spin-current injected into Pt from the (Co/Ni) ferromagnetic reservoir. The more $sp_{Co/Pt}^T$ reduces to a small fraction, the less is the spin-current in Pt leading to a smaller positive AHE. The large specularity in transmission at the Co/Pt interface is then at the origin of the inverted AHE signal one observes in both Type I and type II sample series when $N\simeq 3-10$ is small.

\vspace{0.1in}

We now turn on the discussion of the specularity at (Co/Ni) interfaces (Fig.~\ref{fig:specularity_CoNi}). We consider $sp^T_{Co/Pt}=1$ and $sp^R_{Co/Pt}=0$ for the (Co/Pt) interface corresponding to the optimum conditions for AHE sign inversion. Fig.~\ref{fig:specularity_CoNi}a displays the Anomalous Hall resistance $R_{xy}$ \textit{vs.} $N:1-100$ for different specularity at reflection for each (Co/Ni) interface ($sp_{Co/Ni}^R$) with full transmission. Fig.~\ref{fig:specularity_CoNi}b represents the dependence of $R_{xy}$ on the specularity on transmission at (Co/Ni) ($sp_{Co/Ni}^T$) without reflection, whereas Fig.~\ref{fig:specularity_CoNi}c is the result of joint variation of $sp_{Co/Ni}^R$ and $sp_{Co/Ni}^T=sp_{Co/Ni}^R=sp_{Co/Ni}$ .

Three main remarks can be appended. \textit{i)} For small $N$, one still observes, in any case, a sign inversion of AHE, positive in the case of Pt dominant spin-charge conversion and opposite compared to (Co/Ni). \textit{ii)} the AHE resistance in the region of spin-charge conversion inversion (small $N$) increases when the electron specularity at reflection (Fig.~\ref{fig:specularity_CoNi}a) and at transmission (Fig.~\ref{fig:specularity_CoNi}b) at (Co/Ni) interfaces $sp_{Co/Ni}^{R,T}$ increase from 0 to 1. This indicates a larger spin-current in (Co/Ni), thereby injected into Pt by transport proximity effect. \textit{iii)} On the other hand, the increase of the spin-current in (Co/Ni) volume upon the increase of $sp_{Co/Ni}^{R,T}$ makes that the crossover from positive to negative AHE appears at smaller value of $N$. This should be linked to a larger mean-free path of Co/Ni (Co and Ni) compared to Pt. In the end, a joint increase of the specularity in transmission and in reflection (Fig.~\ref{fig:specularity_CoNi}c) gives rise to the same trends than developed in the case \textit{ii}) and then to  the same conclusions.

\vspace{0.1in}
\emph{Magnetic proximity effects.}
\vspace{0.1in}

\begin{figure}[tbp]
\includegraphics[width=10cm]{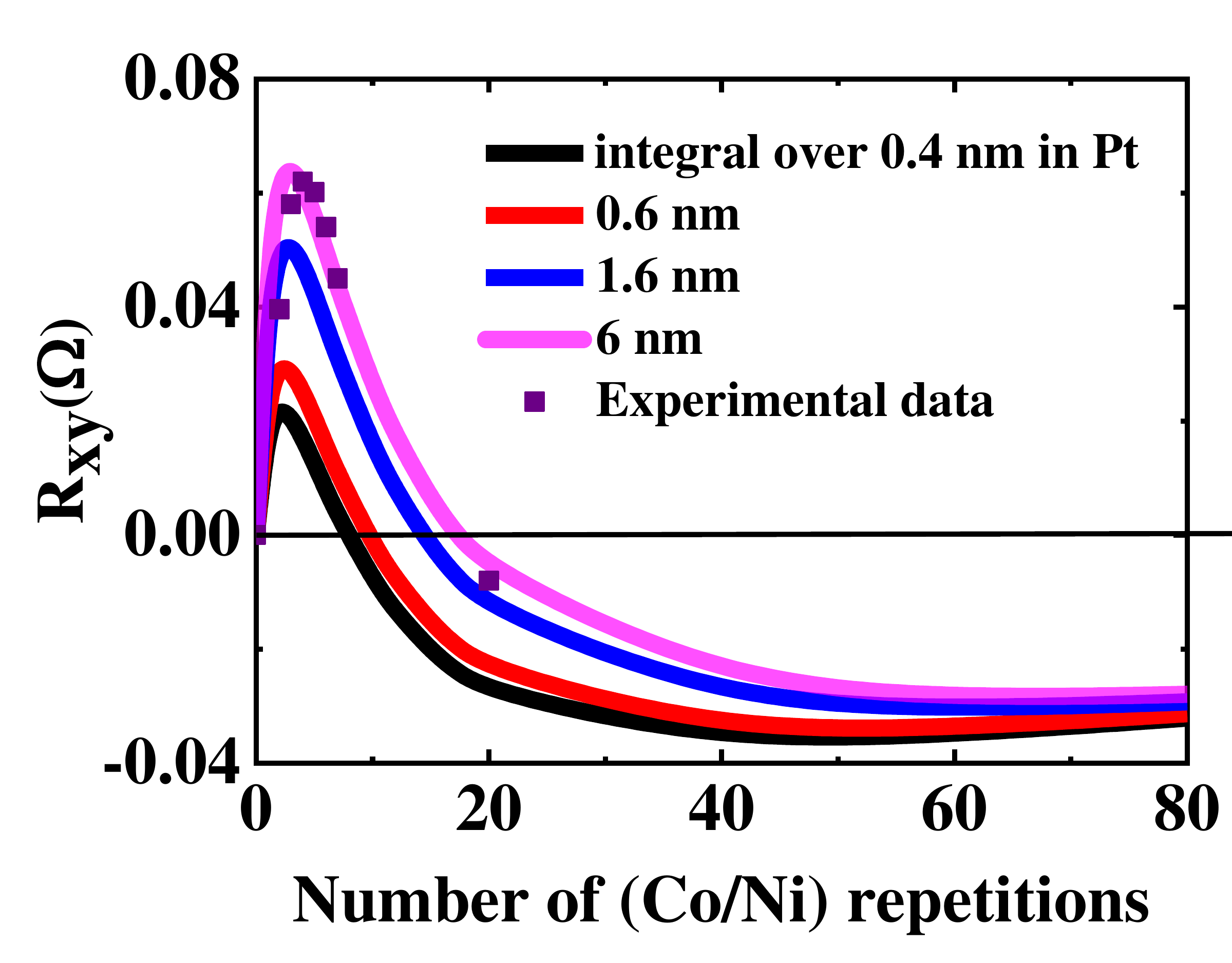}
\caption{Anomalous Hall resistivity \textit{vs.} number of repetitions $N$ of (Co/Ni) in Pt(6)/(Co/Ni)$_N$ \textit{vs.} the a cut-off of the integral of spin-to-charge conversion in the Pt thickness from the (Pt/Co) interface. Squares represent experimental data for type II Pt series.}
\label{fig:integral}
\end{figure}

Moreover, from those simulations, we have performed a different treatment for the integral of the spin-charge conversion in Pt and the results are displayed in the bottom of Fig.~\ref{fig:integral}. We have respectively applied a certain cut-off in the integral up to respectively $z_{cut-off}=0.4, 0.6, 1.6, 6$~nm. This cut-off should mimics the possible role of magnetic proximity effect in Pt leading to induced magnetic moment in Pt over typically 2 or 3 atomic planes. One observes, that using the tabulated parameters in Table~\ref{TableI}, the experimental data may be only recovered when the integral is made on a total thickness of 1.6 or 6~nm well larger than 0.6~nm corresponding to the thickness of 3 atomic planes of Pt. With varying the physical parameters, we were not able to fit the experimental data by using a cut-off of 0.6~nm together with a cross-over occurring at $N=17$, unless considering a spin-Hall angle of Pt larger than $60\%$. This shows that the spin-current proximity effect is the main cause for the  inverted anomalous Hall effect we observe here.

\section{Results of data fitting and conclusions.}

Two different cases may now be distinguished according to the \textit{(i)}~\textit{extrinsic} or \textit{(ii)} and \textit{(iii)} the \textit{intrinsic} nature of the AHE into play. For the case \textit{(ii)} and \textit{(iii)} $\sigma_{xy,i}$ may be expressed as $\sum_s\theta_i^s \sigma_{xx,i}^s$ for both ferromagnetic and normal metals with $\theta_{i}^s$ the local spin Hall angle of layer $(i)$ for the $s$-spin channel~\cite{fert1981}. One has $\theta_i^\downarrow=-\theta_i^\uparrow$ for non-magnetic materials whereas no equivalent relationship exists for a ferromagnet because of the spin-degeneracy lift making $\theta^\uparrow$ and $\theta^\downarrow$ different in absolute value~\cite{fert1981}. However, one may generally assume that $\theta_i^\downarrow$ and $\theta_i^\uparrow$ are of opposite sign. For those calculations, the current density for the $s-$spin channel in the MLs is calculated via the relationship given by the equation~[\ref{js}] where $g^{s}(z,v_{z})$ are the out-of-equilibrium Fermi distributions for spin $s$, solution of the Boltzmann equation  within the MLs, $v_{x,z}$ the Fermi velocity along the current direction ($x$) or along the perpendicular ($z$) to the layers and $S$ the section. $g^{s}(z,v_{z})$ possess two components, one for to the bulk and the other decreasing in $z$ related to the spin-dependent scattering at interfaces that should be found self-consistently. After integration of Eq.~\ref{js}, one has access to $\sigma_{xx,i}^s$ and $\sigma_{xy,i}^s$ respectively. The transverse current is calculated by considering the local transverse conductivity and by summing all contributions.

On the basis of the aforementioned arguments of \textit{extrinsic} \textit{vs.} \textit{intrinsic} SHE mechanism in (Co/Ni) we have proceeded to the four different fitting procedures for AHE in the Pt and Au:W series and retained the best fit. We refer then to Fig.~\ref{fig:3}b for the resulting fits with the different set of parameters given in Table I. In the present case, SML is taken into account in the interfacial scattering matrix at each Pt/Co (with $\delta=0.9$~\cite{jaffres2014,nist2018a,nist2018} or equivalently $p_{sf}=0.6$) and (Co/Ni);(with $\delta=0.25$ or equivalently $p_{sf}=0.3$) interfaces as given in Ref.~\cite{bass}. The fits have been obtained with a SHA for Pt equal to $\theta_{Pt}=\theta_{Pt}^\uparrow=-\theta_{Pt}^\downarrow=+20\pm 2\%$, $\theta_{Au:W_{80}}=+10\pm 1\%$ and $\theta_{Au:W_{130}}=-0.3\pm 0.1\%$ whereas the different models yield (see table I):

\textit{i) intrinsic} SHE mechanism in (Co/Ni) giving $\sigma_{xy}^{int}=-85~S\cdot$cm$^{-1}$;

\textit{ii) extrinsic} SHE mechanism in (Co/Ni) on the majority spin-channel effect giving $\theta _{(Co/Ni)}^\uparrow=-0.9\%$ ($\theta_{(Co/Ni)}^\downarrow=0$) (black fit of Fig.~\ref{fig:3}b);

\textit{iii) extrinsic} SHE mechanism in (Co/Ni) on the minority spin-channel effect giving $\theta_{(Co/Ni)}^\downarrow=-2.2\%$ ($\theta_{(Co/Ni)}^\uparrow=0$) (purple dot fit on Fig.~\ref{fig:3}b).

\textit{iv}) The conductivity for (Co/Ni), $\sigma_{xy}^{int}$, should be compared to the extrinsic one with balanced spin-Hall effect on both spin up and spin down channel $\theta_{(Co/Ni)}^\uparrow=-\theta_{(Co/Ni)}^\downarrow=-1.5\%$ with the corresponding relationship $\sigma_{xy}^{int.}\simeq \theta_{(Co/Ni)}\mathcal{P}\sigma_{xx}$ with $\mathcal{P}\sigma_{xx}=\sigma_{xx}^\uparrow-\sigma_{xx}^\downarrow$ (green fit Fig.~\ref{fig:3}b)

The equivalent spin-conductivity extracted from the \textit{extrinsic} model may be estimated at the vicinity of $-75$~S$\cdot$cm$^{-1}$ for $N=17$, that is close and then in numerical agreement with the case \textit{i}) of \textit{intrinsic} conductivity. The $N=17$ sample corresponds to an equal transverse charge current in Pt and (Co/Ni) giving the condition $\theta_{Pt}\times \Phi_{Pt}^s=\theta_{(Co/Ni)}\times \Phi_{(Co/Ni)}^s$ for $N=17$ where $\Phi^s$ represents the respective fraction of the spin-current in Pt and in (Co/Ni) with $\Phi_{Pt}^s+\Phi_{(Co/Ni)}^s=1$. A value of $\Phi_{Pt}^s=0.07$ as calculated for $N=17$ gives a ratio of about 13 between the spin Hall angles of Pt and (Co/Ni) like extracted from our fit procedure. Fig.~\ref{fig:3}a displays the fits between experiment value of $R_{xy}$ and model for Au:W$_{130}$/(Co/Ni) and Au:W$_{80}$/(Co/Ni) obtained with $\theta_{Au:W_{130}}=-0.3$\% and $\theta _{Au:W_{80}}=$+10\% as experimentally determined in a previous work~\cite{laczkowski2017}. Note that, for both series of samples, at very high number of repetitions $N$ $(N=250)$, $R_{xy}$ of Pt/(Co/Ni) and Au:W/(Co/Ni) merge together towards the intrinsic value of AHE in (Co/Ni), equalling $R_{xy}=-17~m\Omega $. The respective transverse $\rho _{xy}^{\ast}$ in Eq.~\ref{transverse resistance} and longitudinal resistivity $\rho _{xx}^{\ast }$ in Eq.~[2] are also compared to experimental data in Fig.~\ref{fig:2} showing a very good agreement.

Although, one cannot discriminate between the extrinsic \textit{vs.} intrinsic models and simulations for AHE concerning (Co/Ni) for that range of $N=3-40$ and bilayer thicknesses, two major conclusions may be raised. The first one is that, in any case, the consistent spin Hall angle of Pt, + 20~\%, is observed to be clearly enhanced compared to its value extracted from spin pumping-ISHE experiments~\cite{jaffres2014}. No admissible fits may be obtained with typical SHA values for Pt inferior to 20\%.  Such enhancement of the spin-Hall angle in Pt has been already observed in STT-FMR experiments~\cite{nguyen2013,parkin2015,boone2015,nist2018}, lateral spin-valve (LSV) geometry as well as spin-Hall magnetoresistance with Co/Pt~\cite{kawaguchi2018}. This particularly large value of $\theta_{Pt}$ may account for an anisotropy of the electronic scattering time close to the interface. Beyond the change of the intrinsic SHE properties by disorder or energy broadening~\cite{tanaka2007}, electron anisotropic scattering may have for effect to enhance the intrinsic SHA. The second important issue is the magnetic proximity effect: one clearly cannot converge towards a reasonable fit to data when one considers the spin-current integration in space limited to 2-3 Pt atomic planes at the interface with Co unless to consider a value of SHA in Pt of the order of 60-80\%. We consider that magnetic proximity effect of such origin for the AHE inversion we observe cannot play the main role.

\vspace{0.1in}

In conclusion, we evidenced inverted anomalous Hall effect in (Co/Ni) based multilayers grown on thin Pt buffer \textit{via} spin-polarized transport proximity \textit{i.e} spin-current leakage effects. The model and simulations are strongly dependent on the basics electronic transport properties. Using advanced simulation methods for the description of the current and spin-current profiles within multilayers, we have shown that the electronic specularity in reflection and in transmission at inner interfaces and outward surfaces responsible for the increase of resistivity, makes the AHE of thin films and thin multilayers very dependent of the layers quality. We have highlighted an opposite spin Hall angles for Pt and (Co/Ni) and the relevant transport parameters. The sizeable SHA extracted for Pt, +20\%, is opposite to the one of (Co/Ni), giving rise to AHE inversion for thin (Co/Ni) multilayers.  The large SHA of Pt cannot be explained by spin-current proximity effects, and is found to be larger than previously measured in spin pumping-ISHE experiments, effect that may originates from the anisotropy of the electron scattering time in the multilayers. Moreover, we can conclude that AHE data combined to advanced simulation methods may probe main properties of the interfacial spin-orbit interactions in metals.

\vspace{0.1in}

\begin{acknowledgements}
This work was supported by the French Agence Nationale de la Recherche through the ANR Project TOPRISE No. ANR-16-CE24-0017. We  acknowledge financial support from the Horizon 2020 Framework Programme of the European Commission under FET-Open grant agreement no. 863155 (s-Nebula). One of us (T.~H. D) acknowledges Horizon2020 Framework Programme of the European Commission under FET-Proactive Grant agreement no. 824123 (SKYTOP).
\end{acknowledgements}

\newpage

\bibliographystyle{ieeetr}

\end{document}